\documentclass[manuscript]{aastex63}
\usepackage{amsmath}

\begin{document}

\title{Formation and Characteristics of Filament Threads in Double-Dipped Magnetic Flux Tubes}

\correspondingauthor{P. F. Chen} 
\email{chenpf@nju.edu.cn}

\author[0000-0002-4205-5566]{J. H. Guo}
\affiliation{School of Astronomy and Space Science, Nanjing University, Nanjing 210023, China}
\affiliation{Key Laboratory of Modern Astronomy and Astrophysics (Nanjing University), Ministry of Education, Nanjing 210023, China}

\author[0000-0002-4391-393X]{Y. H. Zhou}
\affiliation{Centre for Mathematical Plasma Astrophysics, Department of Mathematics, 
KU Leuven, Celestijnenlaan 200B, B-3001 Leuven, Belgium}

\author[0000-0002-9293-8439]{Y. Guo}
\affiliation{School of Astronomy and Space Science, Nanjing University, Nanjing 210023, China}
\affiliation{Key Laboratory of Modern Astronomy and Astrophysics (Nanjing University), Ministry of Education, Nanjing 210023, China}

\author[0000-0002-9908-291X]{Y. W. Ni}
\affiliation{School of Astronomy and Space Science, Nanjing University, Nanjing 210023, China}
\affiliation{Key Laboratory of Modern Astronomy and Astrophysics (Nanjing University), Ministry of Education, Nanjing 210023, China}

\author[0000-0002-6975-5642]{J. T. Karpen}
\affiliation{Heliophysics Science Division, NASA Goddard Space Flight Center, 8800 Greenbelt Road, Greenbelt, MD 20771, USA}

\author[0000-0002-7289-642X]{P. F. Chen}
\affiliation{School of Astronomy and Space Science, Nanjing University, Nanjing 210023, China}
\affiliation{Key Laboratory of Modern Astronomy and Astrophysics (Nanjing University), Ministry of Education, Nanjing 210023, China}

\begin{abstract}
As one of the main formation mechanisms of solar filament formation, the chromospheric evaporation--coronal condensation model has been confirmed by numerical simulations to explain the formation of filament threads very well in flux tubes with single dips. However, coronal magnetic extrapolations indicated that some magnetic field lines might possess more than one dip. It is expected that the formation process would be significantly different in this case compared to a single-dipped magnetic flux tube. In this paper, based on the evaporation--condensation model, we study filament thread formation in double-dipped magnetic flux tubes by numerical simulations. We find that only with particular combinations of magnetic configuration and heating, e.g., concentrated localized heating and a long magnetic flux tube with deep dips, can two threads form and persist in a double-dipped magnetic flux tube. Comparing our parametric survey with observations, we conclude that such magnetically connected threads due to multiple dips are more likely to exist in quiescent filaments than in active-region filaments. Moreover, we find that these threads are usually shorter than independently trapped threads, which might be one of the reasons why quiescent filaments have short threads. These characteristics of magnetically connected threads could also explain barbs and vertical threads in quiescent filaments.
\end{abstract}


\section{Introduction} \label{sec:intro}

Solar filaments are composed of cold, dense plasma suspended in the hot corona. Their typical temperature is two orders of magnitude lower but their density is two orders of magnitude higher than their surroundings \citep{Labrosse2010,Mackay2010}. They appear as elongated dark structures against the solar disk and as bright cloud-like structures above the solar limb where they are called prominences. Filaments are generally formed along filament channels, which follow along the polarity inversion line (PIL). While magnetic dips may not be necessary in some filaments where continuous formation and drainage of cold threads maintain the dynamic appearance of filaments \citep{Karpen2001, Zou2016, Zou2017}, it is believed that most filaments are supported against gravity by the Lorentz force in magnetic dips, where the cold plasma is trapped. Generally, magnetic structures with dips can be described by two idealized models: sheared arcades \citep{Kippenhahn1957, Antiochos1994}, and magnetic flux ropes \citep{Kuperus1974}. Magnetic dips can exist naturally in quadrupolar magnetic configurations. Sheared-arcade filament channels can be formed either by relative displacements of the magnetic footpoints (shear) on either side of the PIL or by helicity condensation \citep{Antiochos2013}. To obtain flux ropes whose axes reside above the photosphere, two major mechanisms have been proposed: a flux rope emerges directly from the convection zone \citep{Fan2001, Okamoto2008}, or a flux rope is formed through magnetic reconnection between the legs of a strongly sheared arcade \citep{Ballehooijen1989} or between adjacent sheared arcades \citep{Mackay2008, Xia2014b}. Statistics of erupting filaments from 2010 May to 2015 December showed that $89 \%$ of filaments are inverse-polarity filaments with a flux-rope configuration, while only $11 \%$ are normal-polarity filaments with a sheared-arcade configuration \citep{Ouyang2017}, based on the drainage method \citep{Chen2014}. 

Once the magnetic structure of a filament channel is formed, an ensuing important issue is how the cold plasmas accumulate in the corona. By analyzing the spectroheliograms in the \emph{Skylab} plate collection, \citet{Spicer1998} found that the Ne and Mg abundance ratio in filaments deviates from that of the corona but is similar to that of the chromosphere, indicating that the filament material might originate from the lower atmosphere, and then somehow is transported to the corona. Based on the observations and simulations, three models have been proposed: the injection model \citep{An1988, Wang1999, Wang2019}, the levitation model \citep{Lites2005, Okamoto2008, Zhao2017}, and the evaporation--condensation model \citep{Antiochos1999}. In the injection model, cold chromospheric plasma is directly ejected into the corona by chromospheric magnetic reconnection. Note that most of the observational evidence for the injection model comes from active-region filaments, and this model still cannot explain the formation of high, long quiescent filaments. In the levitation model, cold material is lifted by a rising magnetic flux rope and then resides in the magnetic dips. The evaporation--condensation model has attracted more attention, due to decades of modeling efforts and to its ability to explain the sudden appearance of many filaments without visible cool material rising from the chromosphere. In this model, the chromosphere is a mass reservoir, and chromospheric materials are heated by localized heating and evaporated into the corona, where they condense due to catastrophic cooling. This model has been extensively investigated via numerical simulations \citep{Antiochos1999, Karpen2003, Karpen2005, Karpen2006, Karpen2008, Xia2011, Xia2012, Xia2014, Xia2016, Zhou2014, Viall2020}. \citet{Zhou2020} also proposed that turbulent heating in the chromosphere can successfully explain both the origin of filament threads and their counterstreaming flows. Recently, \citet{Huang2021} made an effort to unify the direct injection model and the evaporation--condensation model.

All of the previous numerical simulations were based on the assumption that there is only a single dip along each magnetic field line. However, both simulations and data-based extrapolations indicate that there might be more than one dip on some field lines within filament channels. For example, in the coronal magnetic field constructed by \citet{Jing2010} through nonlinear force-free field (NLFFF) extrapolation, two dips are present along some magnetic field lines. Recently, \citet{Guo2019} reconstructed a flux-rope configuration for an observed filament with the regularized Biot-Savart laws (RBSL), and found that the average twist is larger than two turns. In the head-to-tail model presented by \citet{Martens2001} and modeled more comprehensively by \citet{Devore2005} for the filament formation and evolution, two magnetic dips along one field line is a natural result. \citet{Mackay2009} studied the magnetic structure evolution when a small magnetic element approaches a solar filament, and found that some magnetic field lines have double dips even when the total twist is less than two turns. Other magnetic modeling works also showed that some quiescent or polar crown prominences could be supported by single, highly-twisted flux ropes \citep{Su2012, Su2015, Mackay2020}. Many prominences outside active regions appear to form in multiple segments, similar to the schematic picture of Figure 6 in \citet{Chen2011}. These segments can be joined through reconnection, yielding occasional field lines with two dips \citep{Devore2005}. If two dips exist along one field line, both formation and subsequent dynamics of threads in the two dips might be very different from the well-studied single dip case. 

The research described here uses one-dimensional (1D) hydrodynamic simulations with thermal conduction and optically thin radiation to explore how multiple magnetic dips affects filament thread formation. This paper is organized as follows. The numerical method is described in Section \ref{sec:met}. Results are presented in Section \ref{sec:res}, which are followed by discussions in Section \ref{sec:dis} before a summary in Section \ref{sec:sum}. 

\section{Numerical setup} \label{sec:met}

For simplicity, we consider the magnetic field line as a rigid tube. With this approximation, similar to our previous works \citep{Xia2011, Zhang2013, Zhang2020, Zhou2014}, we numerically simulate filament mass formation with the following one-dimensional (1D) radiative hydrodynamic equations,

\begin{eqnarray}
 && \frac{\partial \rho}{\partial t} +\frac{\partial (\rho v)}{\partial s}=0, \label{eq1}\\
 && \frac{\partial (\rho v)}{\partial t}+\frac{\partial (\rho v^2+p)}{\partial s}=\rho g_\parallel(s), \label{eq2}\\
 && \frac{\partial \varepsilon}{\partial t}+\frac{\partial ( \varepsilon v+pv)}{\partial s} =\rho g_\parallel v+H(s,t)-n_{e}n_{H}\Lambda(T)+ \frac{\partial}{\partial s}(\kappa \frac{\partial T}{\partial s}), \label{eq3}
\end{eqnarray}
where $s$ is the distance along the magnetic flux tube, $g_\parallel$ is the field-aligned gravity component, $\varepsilon =\frac{\rho v^2}{2}+\frac{p}{\gamma -1}$ is the total energy density, $\gamma =\frac{5}{3}$ is the adiabatic index, $\kappa =10^{-6}\ T^{\frac{5}{2}}\ \rm erg\ cm^{-1}\ s^{-1}\ K^{-1}$ is the Spitzer heat conductivity, $H(s,t)$ is the heating rate, $\Lambda(T)$ is the radiative loss function, and others have their general meanings. We presume that the plasma is fully ionized and the helium abundance is 0.1, so that the mass density is $\rho =1.4\ m_{p}n_{H}$, and the gas pressure is $p=2.3\ n_{H}k_{B}T$. Energy Equation (\ref{eq3}) includes the heating $H(s,t)$, the optically thin radiative term $n_{e}n_{H} \Lambda(T)$, and the heat conduction $\frac{\partial}{\partial s}(\kappa \frac{\partial T}{\partial s})$. $H(s,t)$ is divided into the steady background heating $H_0(s)$ and the finite-duration localized heating $H_{_{l}}(s,t)$, where the background heating is used to maintain the hot corona, and the localized heating is applied to drive chromospheric evaporation. Similar to \citet{Xia2011}, the optically-thin radiative loss function $\Lambda(T)$ is truncated at 16000 K. As the temperature falls below 16000 K, the optically thin radiation approximation is no longer valid, and radiative transfer should be considered if the detailed structures of filaments are to be considered, which is beyond the scope of this paper.

Regarding the background heating, a thermal-energy input profile that drops  exponentially with height has been shown to agree with the observed thermal structure in the corona \citep{Klimchuk2010}. Hence, the background heating $H_{0}$ is described by Equation (\ref{eq4}), where $E_0$ is the heating amplitude, $H_{m}=L/2$ \citep{Withbroe1988} is the scale length, and $L$ is the total length of the flux tube: 

\begin{eqnarray}
 && H_{0}(s) = \begin{cases} E_{0}\exp(-s/ H_{m} )  & s<L/2,\\ E_{0}\exp[-(L-s)/ H_{m}] & L/2 \leq s <L. \end{cases} \label{eq4}
 \end{eqnarray}

To drive chromospheric evaporation, we set the localized heating function as Equation (\ref{eq5}),

 \begin{eqnarray}
 && H_{_{l} }(s) = \begin{cases} E_{1}  & s  \leq  s_{_{tr} }, \\ E_{1}\exp[-(s-s_{_{tr} })/ \lambda]  &  s_{_{tr} }<s \leq L/2, \\ fE_{1}\exp[-(L-s_{_{tr} }-s)/ \lambda] & L/2<s \leq L- s_{_{tr} }, \\ fE_{1}  & s > L-s_{_{tr} }, \end{cases}   \label{eq5}
\end{eqnarray}
$E_{1}= 10^{-2}$ $\rm erg\,cm^{-3}\,s^{-1}$ the localized heating amplitude, $s_{_{tr}}$ the height of the transition region, $\lambda$ the scale length and $f$ the ratio of the localized heating amplitudes between the two footpoints. The localized heating is ramped up linearly over a duration $\tau_{_{l}}$ and then shut off quickly. This finite-duration localized heating may be due to the nanoflare storm \citep{Parker1988, Klimchuk2015}, and the heating is effectively steady if the interval between storms is less than the coronal radiative cooling time \citep{Karpen2008}. In this scenario, the heating may stop when the slipping reconnection velocity during the nanoflare \citep[non-ideal velocity,][]{Yang2018} is equal to zero. Transient chromospheric reconnection with finite lifetime is also a good candidate for the localized footpoint heating.

The field-aligned gravity component $g_{\parallel}$ is determined by the geometric configuration of the magnetic flux tube, $g_{\parallel}(s)=\ \boldsymbol{g_{\odot}} \cdot \hat{e_{s}}$, where $\boldsymbol{g_{\odot}}=-274\hat{e_{z}} \ \rm m \ s^{-2}$, and $\hat{e_{s}}$ is the unit tangential vector along the field line. In this paper, two different types of magnetic configurations are considered. First, we construct a simple helical flux tube with two turns, and hence two dips, as used by \citet{Zhou2017}. Then, in order to consider more complex magnetic configurations, we also construct a three-dimensional magnetic flux rope using the Titov-D\'emoulin-modified (TDm) model \citep{Titov2014, Titov2018}, and select some field lines to perform our 1D simulations.

\begin{figure}[ht!]
\centering
\includegraphics[scale=0.9]{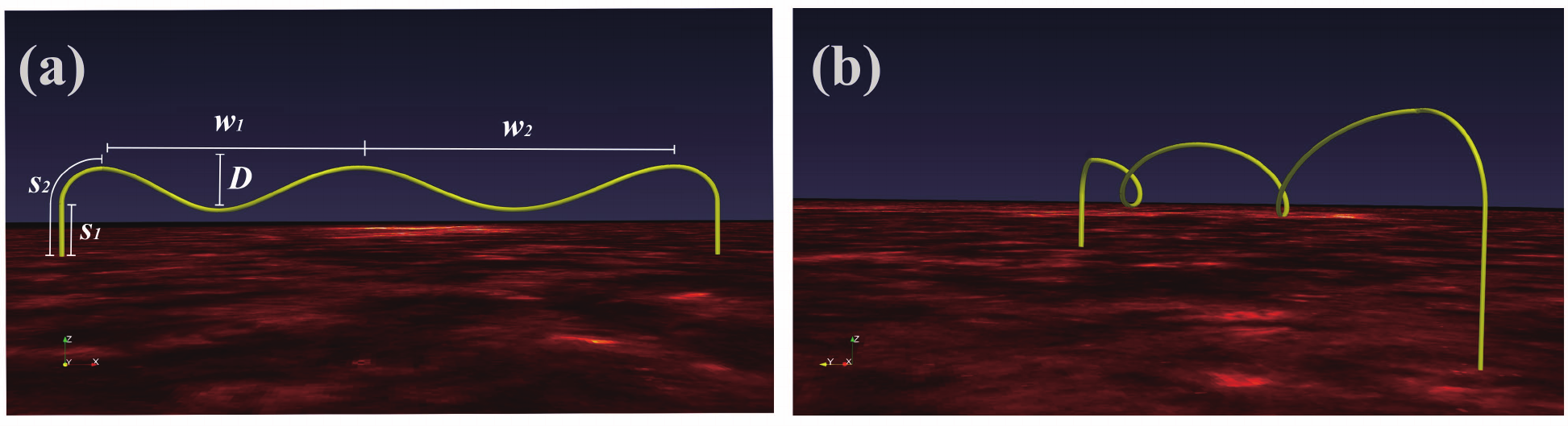}
\caption{Geometry of a helical flux tube used for one-dimensional hydrodynamic simulations of filament threads. (a) and (b) are the front and side views, respectively.}
\label{fig1}
\end{figure}

The helical field line consists of three portions: a vertical leg $0\leq s \leq s_{1}=15\ \rm Mm$, a quarter-circular arc $s_1\leq s\leq s_2=22.85\ \rm Mm$, and a central helical part, as shown in Figure \ref{fig1}. The central helical part is described by the following formulae,
\begin{eqnarray}
 && \begin{cases} x=(\frac{\theta}{4\pi})^{n}l;  \\ y=0.5D\sin \theta; \\ z=0.5D\cos \theta;\end{cases}  
\label{eq6}
\end{eqnarray}
where $x$ is the horizontal coordinate along the main axis of the helix, $y$ is the horizontal coordinate perpendicular to the $x$-axis, the $z$-axis is vertical, and $\theta$ ranges from 0 to $4 \pi$. The helix structures are controlled by the dip depth ($D$), the length of the main axis ($l$), and the asymmetry metric ($n$). 

\begin{figure}[ht!]
\centering
\includegraphics[scale=0.2]{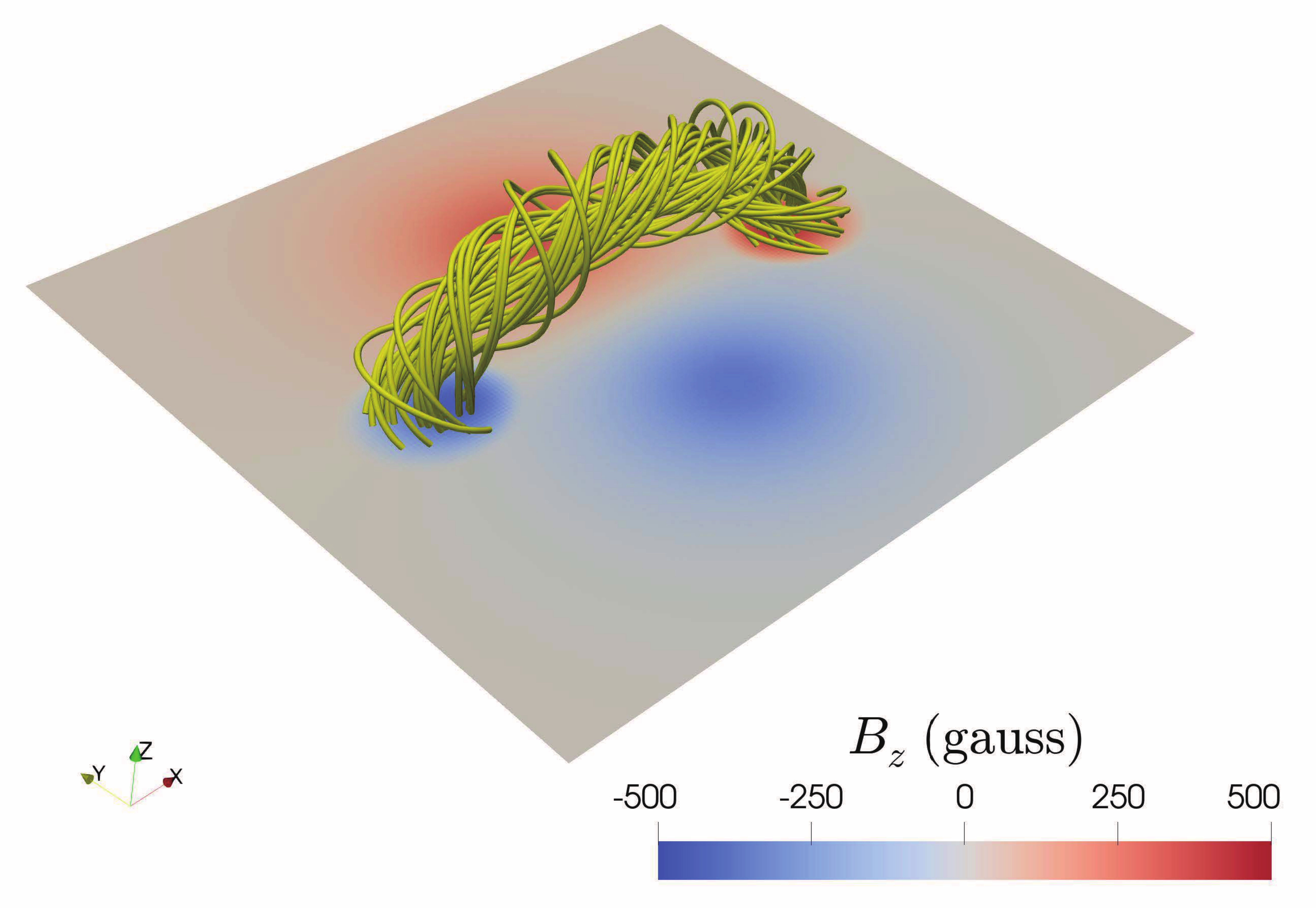}
\caption{A magnetic flux rope constructed by the TDm model. The average twist of the flux rope is about 1.83 turns. Selected field lines from this flux rope were used for 1D simulations of filament threads. }
\label{fig2}
\end{figure}

Following \citet{Titov2018}, we construct a flux rope that is embedded in a bipolar background field. First, we calculate the background magnetic field $\boldsymbol{B_{\rm q}}$ with two sub-photosphere magnetic charges of strength $q$, which lie on the symmetry axis under the photosphere at the depth $d_{q}$. The two charges are separated by $2L_{q}$. Second, we calculate the physical parameters for the RBSL method. We need to set two geometric parameters of the flux rope: the axis path and the minor radius $a$. The axis path has a semicircle shape that follows a contour line of the external field $B_{\perp}$ in the plane perpendicular to the symmetry axis. Since the RBSL method requires a closed path, this is done by adding the sub-photosphere semicircle arc. Thus, the full path is a circle with a radius $R_{c}$. Then, the current $I$ and magnetic flux $F$ are determined by Equation (7) in \citet{Titov2014} and Equation (12) in \citet{Titov2018}, respectively. Third, we construct the flux rope with the aforementioned parameters and embed it into the background field. Finally, we relax the combined magnetic field by the magneto-frictional method \citep{Guo2016a, Guo2016b}. In this paper, we choose the following parameters: $q=100 \ \rm T \ Mm^{2}$, $L_{q}=50\rm \ Mm$, $d_{q}=50\rm \ Mm$, $R_{c}=94.22\rm \ Mm$, and $a=25 \rm \ Mm$. After relaxation, the force-free metric $\sigma_{J}=0.0892$, and the divergence-free metric $<|f_{i}|>=1.34 \times 10^{-5}$ \citep[see][for details of the two metrics]{Guo2016b}. These two values are small enough to guarantee the model to be in a force-free and divergence-free state. The final magnetic-field distribution is illustrated in Figure \ref{fig2}. The twist of a flux rope and the number of dips are highly correlated, so we also calculate the twist using the formula provided by \citet{Berger2006}. In this case, the mean twist is about 1.83 turns. We selected some magnetic field lines from this flux rope for our simulations.

\begin{figure}[ht!]
\centering
\includegraphics[scale=0.9]{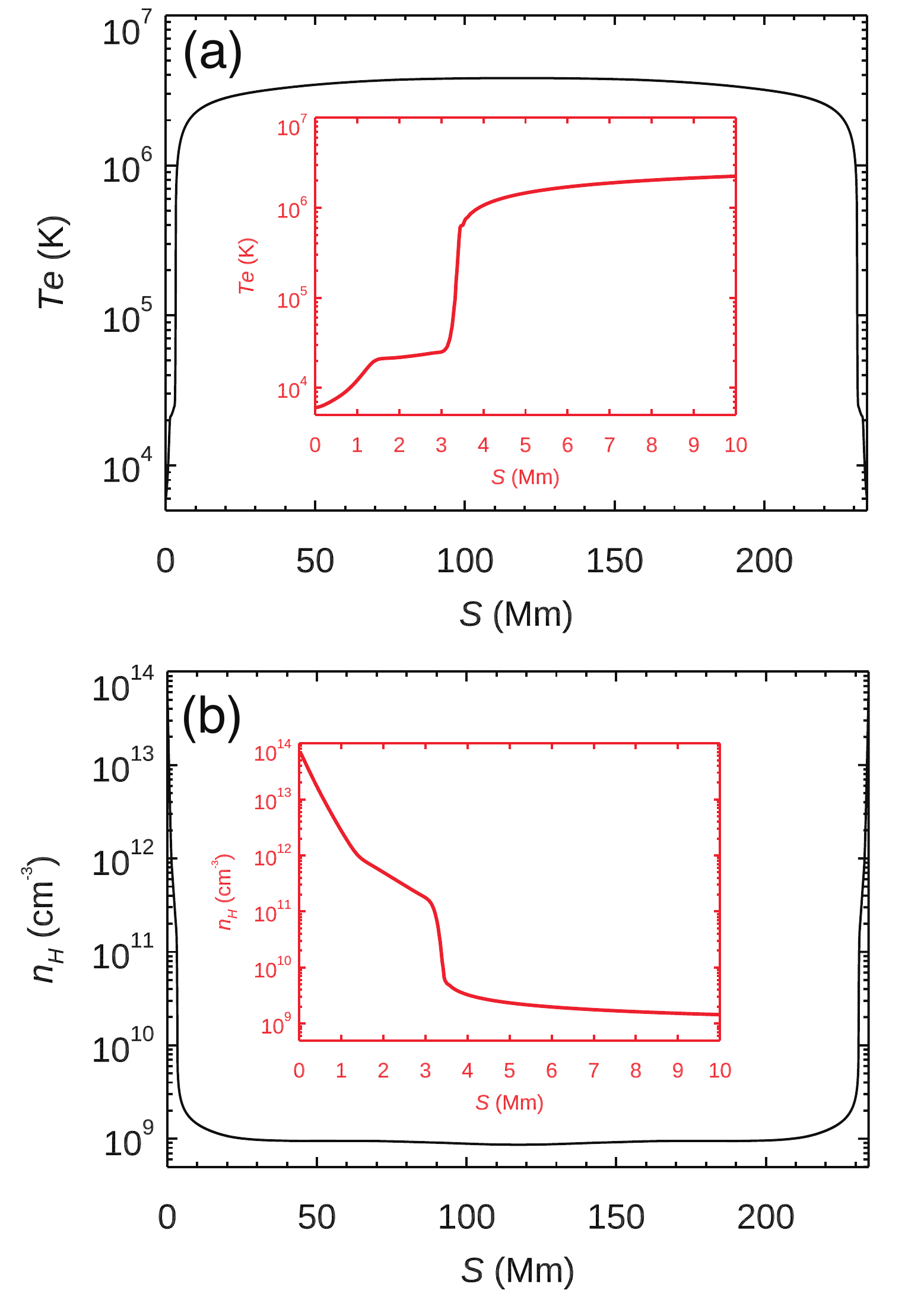}
\caption{The temperature (a) and number density (b) distributions along the flux tube in the initial state. The insert figures show details of the temperature and number density profiles below 10 Mm.}
\label{fig3}
\end{figure}

To obtain a hydrostatic quiet-Sun atmospheric model, similar to our former simulations \citep{Zhou2017}, we first assume a temperature distribution that is a function of height ($z$) governed by the following formula,
 \begin{eqnarray}
  && T(z)=T_{1}+ \frac{1}{2}(T_{2}-T_{1})(\tanh( \frac{z-h_{0}}{w_{0}})+1),  \label{eq7}
 \end{eqnarray}
where $T_1=$ 6000 K is the chromospheric temperature, $T_2=$ 1 MK is the coronal temperature, $h_0=$ 2.72 Mm sets the height of the midpoint of the transition region, and $w_0=$ 0.25 Mm controls the thickness of the transition region. Thus, the thickness of the chromosphere in the initial state is about 3 Mm. Then the density distribution is determined by the hydrostatic equilibrium \citep[see][for details]{Xia2011}. However, only the force equilibrium is satisfied in this stage but the energy equilibrium is not. Hence, we relax the initial state to a stable energy equilibrium state with the background heating only ($\sim$ 57 min). Figures \ref{fig3}a and \ref{fig3}b show the initial temperature and number density distributions, respectively, along the flux tube after the relaxation.

The 1D radiative hydrodynamic equations are numerically solved with the Message Passing Interface Adaptive Mesh Refinement Versatile Advection Code \citep[MPI-AMRVAC\footnote{http://amrvac.org},][]{Xia2018, Keppens2020}. We use six levels of adaptive mesh refinement, with 960 base-level grids, which leads to an effective resolution ranging from 5.81 to 13.40 km per grid point. Since the influence of the corona on the lower atmosphere is negligible, the boundary conditions are fixed at the two footpoints of each modeled magnetic field line. The density in the ghost cells is set according to the hydrostatic equilibrium condition, to avoid downflows at the boundary.

\section{Numerical Results}\label{sec:res}
\subsection{Helical flux tube} \label{sec:helical}
\subsubsection{Heating characteristics} \label{sec:helical_h}
\begin{figure}[ht!]
\centering
\includegraphics[scale=0.65]{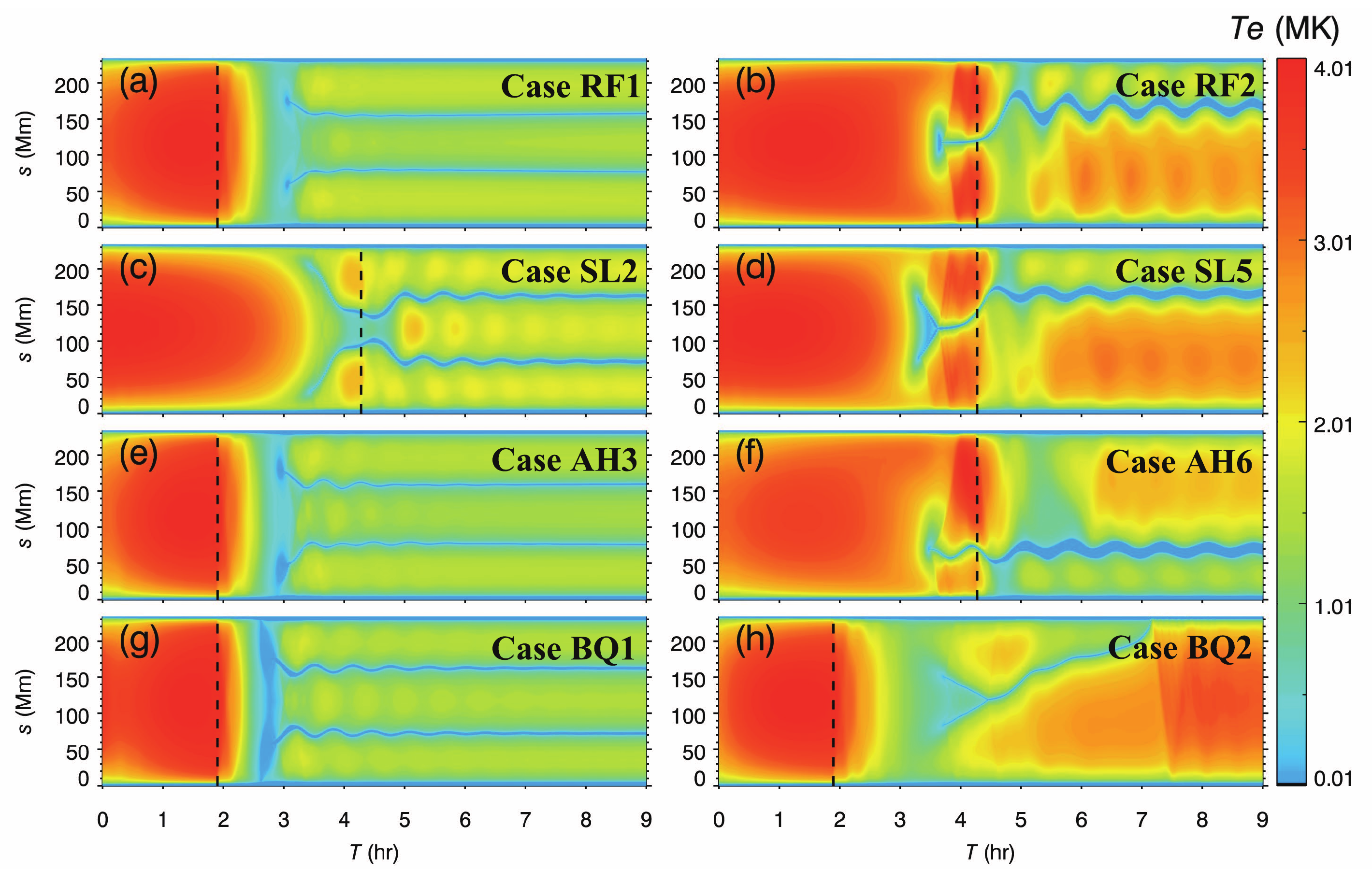}
\caption{The evolution of the temperature distribution along the flux tube in 8 cases with different heating parameters. Four rows from top to bottom show the influences of the localized heating time $\tau_{_{l} }$, the localized heating scale length $\lambda$, the ratio of the localized heating rate between two footpoints $f$, and the amplitude of background heating $E_{0}$ on the formation of two threads, respectively. Panels (a)--(h) represent the results in cases RF1, RF2, SL2, SL5, AH3, AH6, BQ1, and BQ2, respectively. The black dashed lines represent the time when the localized heating is halted.}
\label{fig4}
\end{figure}

\begin{figure}[ht!]
\centering
\includegraphics[scale=0.4]{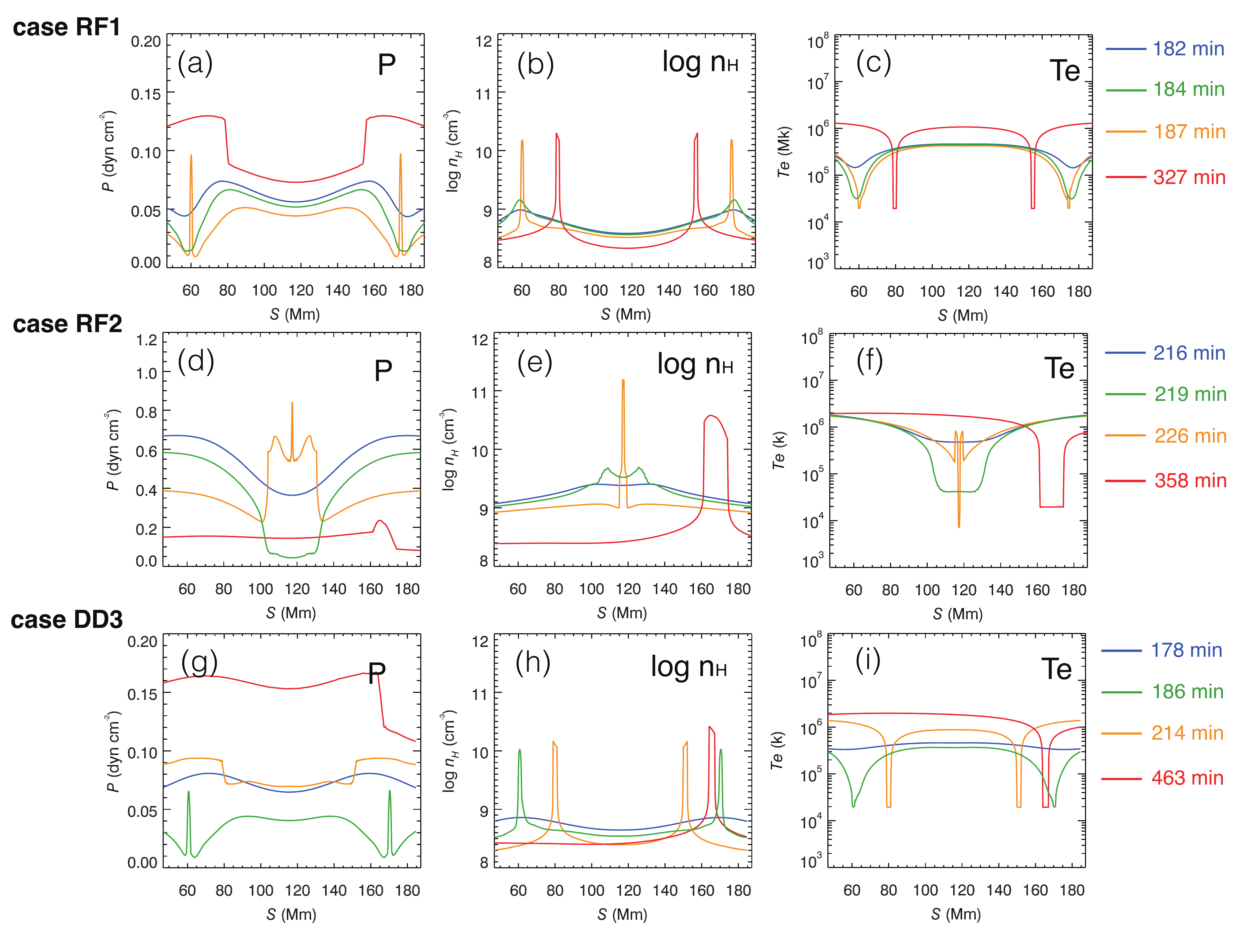}
\caption{The distribution of the pressure (a, d, g), number density (b, e, h) and temperature (c, f, i) along the flux tube in cases RF1 (top panels), RF2 (middle panels) and DD3 (bottom panels) at different times during each simulation. The line colors denote different times.}
\label{fig5}
\end{figure}

Previous simulations have demonstrated that the localized heating is a crucial component of the thread formation process in singly-dipped flux tubes \citep{Antiochos1999, Karpen2006, Karpen2008, Xia2011}. Here we determine whether the localized heating can also produce two threads. In the following simulations, we keep the geometry of the flux tube unchanged, namely, $D=9 \ \rm Mm$, $l=180 \ \rm Mm$, $n=1.0$, and adjust the background heating, $E_{0}$, the localized heating duration, $\tau_{_{l} }$, the localized heating scale length, $\lambda$, and the ratio of the localized heating rate between two footpoints, $f$. In order to make comparisons, we set a benchmark case RF1 with $\tau_{_{l} }=115$ min, $\lambda=10\ \rm Mm$, $f=1.0$ (symmetric heating), and $E_{0}=2.0\times \rm 10^{-4} \ erg\ cm^{-3}\ s^{-1}$. Figure \ref{fig4}a depicts the temporal evolution of the temperature distribution along the flux tube in case RF1. Evidently, two threads are formed and remain in the two magnetic dips after $t$=3 hr. The pressure, number density, and temperature distributions along the flux tube in case RF1 before and after the coronal condensation are revealed in Figures \ref{fig5}a--c, respectively. At $t= 3.067$ hr, the temperature and pressure at $s=60\ \rm Mm$ and $s=175\ \rm Mm$ (away but not far from the dip centers) decreases drastically, sucking the surrounding coronal plasma toward these cooling regions, which leads to a catastrophic rise in density. Soon the two cold filament threads shift toward each other and are trapped in dips, and a low pressure region is formed between the two threads (Figure \ref{fig5}a). 

\begin{figure}[ht!]
\centering
\includegraphics[scale=0.45]{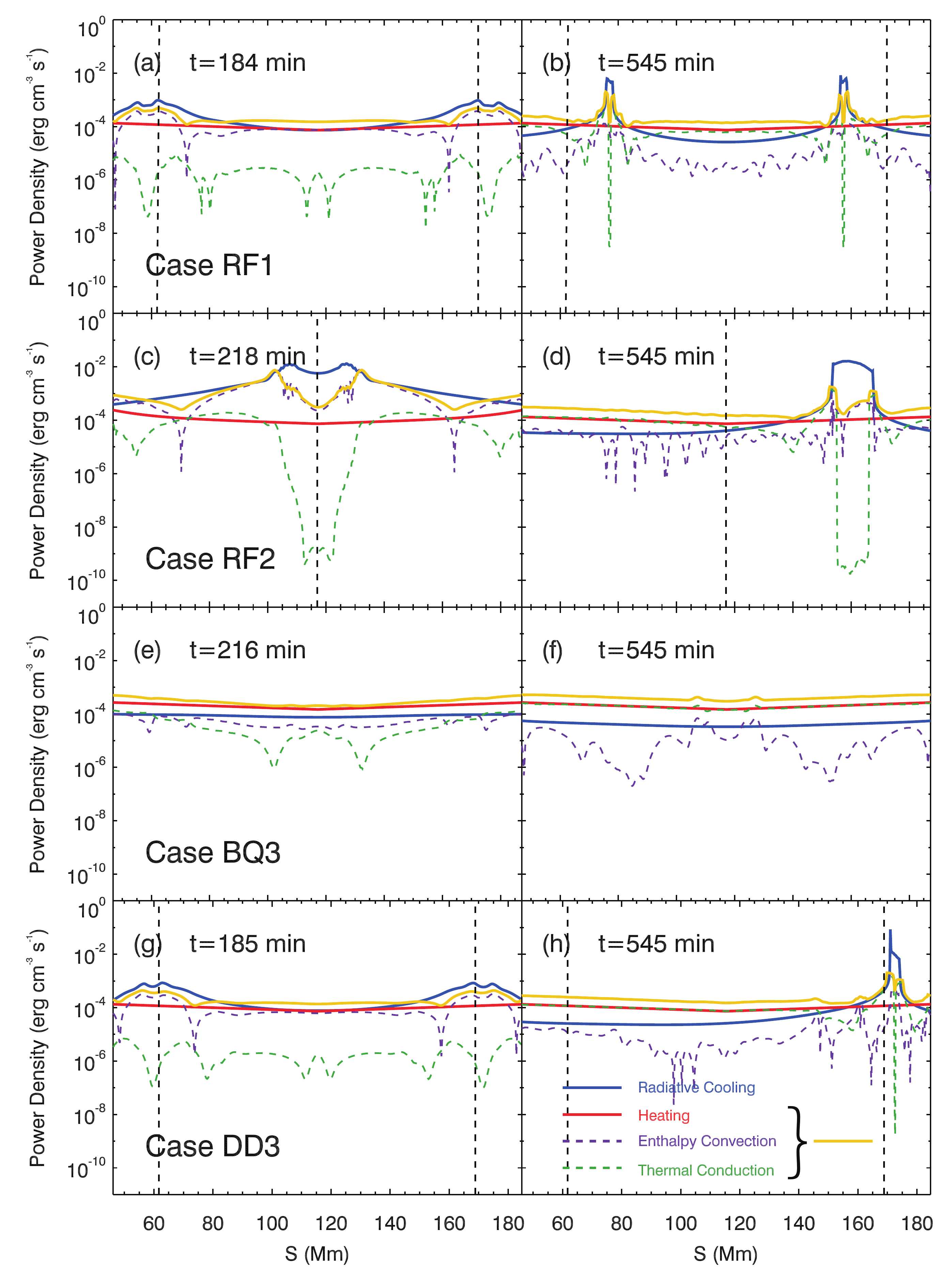}
\caption{Absolute value distributions of different energy terms in (a, b) case RF1; (c, d) case RF2; (e, f) case BQ3; (g, h) case DD3, including radiative cooling (blue solid lines), heating (red solid lines), thermal conduction (green dashed lines), and enthalpy convection (purple dashed lines) of the coronal portion at the onset times of the catastrophic cooling and the end of simulations. Yellow lines represent the sum of the heating, thermal conduction and enthalpy convection. Vertical dashed lines represent the locations of thread formation. Note that thermal runaway cannot occur in case BQ3, as shown in panels (e) and (f).}
\label{fig6}
\end{figure}

To understand the contributions of different energy sources to the coronal structure with filament threads, in Figure \ref{fig6} we show the absolute value distributions of the energy terms, including heating ($H=H_{0}+H_{_{l}}$), radiative losses ($n_{e}n_{H}\Lambda(T)$), thermal conduction ($\frac{\partial}{\partial s}(\kappa \frac{\partial T}{\partial s})$) and enthalpy convection ($\frac{\partial}{\partial s}(\frac{\gamma pv}{\gamma-1})$). Figure \ref{fig6}a illustrates that the catastrophic coolings occur where the radiative losses exceed the sum of the heating, conductive flux, and enthalpy flux, which is consistent with the thermal non-equilibrium (TNE) theory \citep{Antiochos1999, Klimchuk2019, Antolin2020}. In addition, we find that the radiative cooling is dominant in the thread portion, and thermal conduction is important only at the condensation boundaries.

\begin{figure}[ht!]
\centering
\includegraphics[scale=1.0]{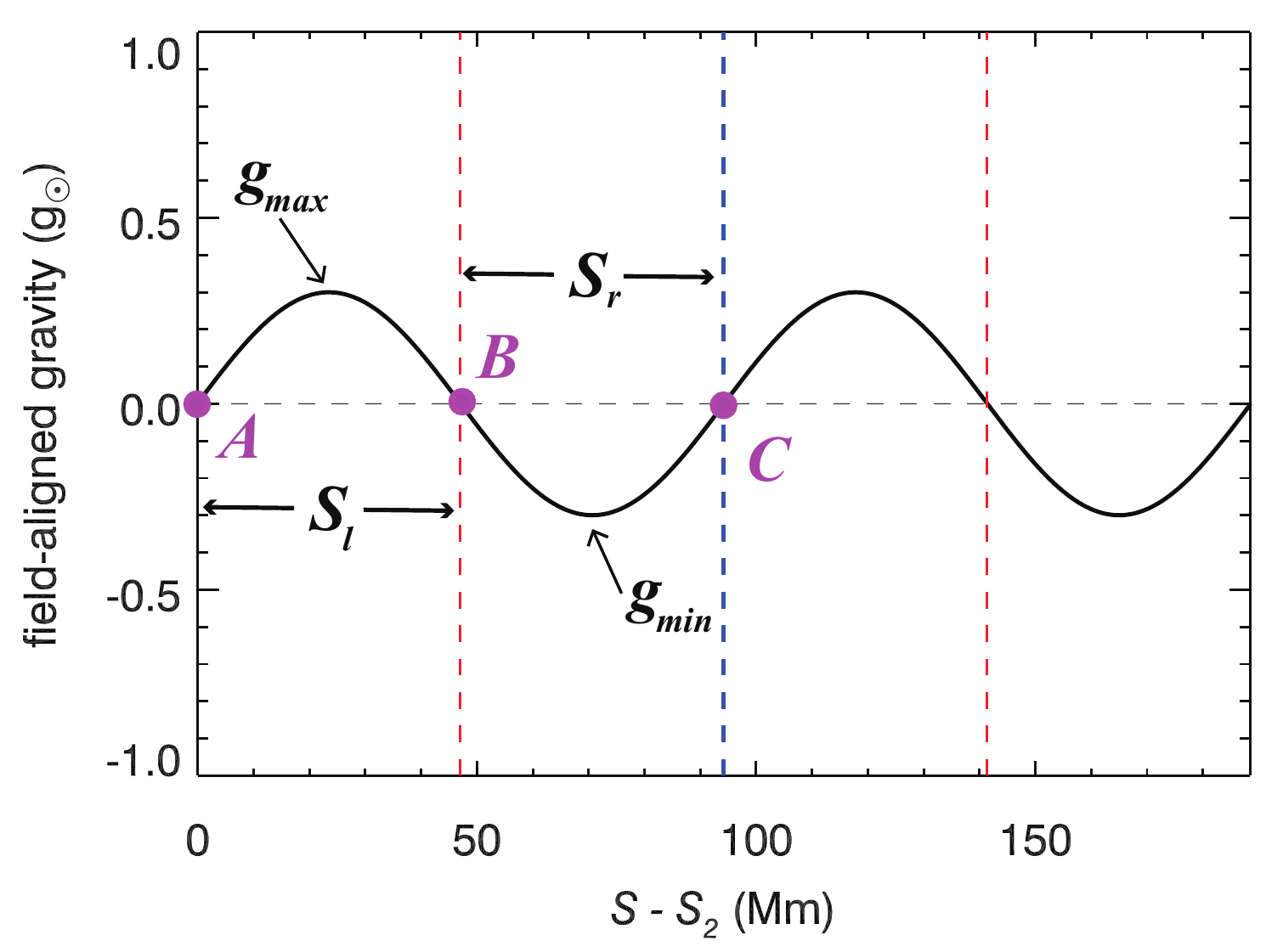}
\caption{The field-aligned gravity profile of the helical section of the flux tube in case RF1, where point A (C) represents the lefthand (righthand) dip shoulder, and point B denotes the dip center, $S_{l}$ ($S_{r}$) is the length of left (right) portion, $S_{_{d}}=S_{l}+S_{r}$ is the total length, and $g_{_{max}}$ ($g_{_{min}}$) is the maximum (minimum) field-aligned normalized gravity inside the dip. The vertical dashed lines marked in red (blue) denote the locations of dip center (shoulder). Note that point C is also the midpoint of the entire flux tube.}
\label{fig7}
\end{figure}

Table \ref{tab1} lists the input parameters and the simulation results for all cases in our parameter study of heating characteristics, where the simulation results include the number of threads, onset time of the catastrophic cooling, and the thread length. We perform five simulations with localized heating duration $\tau_{_{l} }=$ 143, 172, 201, 229, and 258 min, as listed in Table \ref{tab1}. When $\tau_{_{l} } > \tau_{_{form} }$ ($\tau_{_{form}}$ is the onset time of the catastrophic cooling), we find that only one thread is generated even in a double-dipped flux tube. One example is case RF2, whose temperature evolution is illustrated in Figure \ref{fig4}b, and the pressure, number density, and temperature distributions along the flux tube are shown in Figures \ref{fig5}d--\ref{fig5}f. In case RF1, the condensations are formed on the outer downslope of the dips (i.e., regions between points A and B in Figure \ref{fig7}) after the entire coronal portion of the flux tube cools, and then persist in dips. In case RF2, the strong localized footpoint heating continues past the thread formation, which suppresses the cooling of the entire flux tube in the coronal portion. Thus, one condensation forms at the midpoint of the flux tube and then falls into one dip. This thermodynamic process is similar to single-dipped cases with symmetric heating \citep{Karpen2001, Xia2011}. Compared to the magnetically connected threads in case RF1, the independently trapped thread in case RF2 is denser (Figure \ref{fig5}e) and has larger peak radiative losses than the magnetically connected threads in case RF1 (Figure \ref{fig6}d).

Second, we investigate the effect of the localized heating scale length $\lambda$. Previous studies of single-dipped field lines have demonstrated that, if the localized heating scale length $\lambda$ is small enough, two condensations will form on the two shoulders of the flux tube. Depending on whether the condensations form inside or outside the dip, the condensations can either approach and coalesce or fall to the nearest footpoint \citep{Karpen2006, Xia2011}. Accordingly, we expect that decreasing $\lambda$ to some critical value would generate two threads. We take case RF2 as a benchmark rather than case RF1, because only one thread forms in case RF2; hence, the effect of varying $\lambda$ can be easily recognized. Keeping other parameters the same as in case RF2, we perform five simulations with $\lambda$ = 3, 4, 5, 7, and 9 Mm. It is found that two threads can exist only when $\lambda =4\ \rm Mm$ (Figure \ref{fig4}c). In case SL1, two condensations form and then fall into the chromosphere. In other cases (cases SL3--SL5), although coronal condensations start to form in two dips, the inward pressure-gradient force drives the two segments to approach each other and coalesce at the midpoint of the flux tube. The coalesced thread finally falls to one dip, as in cases SL5 (Figure \ref{fig4}d). Thus, localized heating with scales below 10 Mm can indeed drive catastrophic cooling to form two condensations, but only a subset of $\lambda$ values ($\lambda/L \sim 1/59$) yields two stable threads that persist. In our results, the threshold of the localized heating scale ratio from one condensation (large $\lambda/L$) to two condensations (small $\lambda/L$) is 1/23, which is slightly larger than 1/28 \citep{Xia2011} and is between 1/20 and 1/33 \citep{Muller2004}.

The observed magnetic-field distribution in the lower atmosphere is extremely non-uniform, particularly in the vicinity of filament channels, which implies that the heating characteristics of two footpoints might be asymmetric \citep{Karpen2005, Mok2005, Mok2008, Chen2014}. To study the effects of such an asymmetry, we perform simulations with different localized heating ratios $f=1.1,\ 1.4$ and 1.7, and different durations $\tau_{_{l}}$=115 and 258 min, as listed in Table \ref{tab1}. Comparing case AH3 (Figure \ref{fig4}e) and case RF1 (Figure \ref{fig4}a), we infer that when the localized heating duration is short the heating asymmetry has little effect on the number of threads. In cases AH1--AH3, the catastrophic cooling takes place after the localized heating is turned off, so the asymmetry of the localized heating cannot regulate the dynamics of threads. However, when the localized heating endures past the commencement of coronal condensation, the condensation tends to form and remain in the dip near the less heated footpoint rather than at the loop center (Figure \ref{fig4}f), consistent with the single-dip results of \citet{Karpen2003} and \citet{Xia2011}. Basically, the asymmetric localized heating produces a pressure imbalance along the flux tube, placing the stagnation point of the evaporation flows closer to the more weakly heated footpoint and driving the thread toward that footpoint. 

The above results indicate that, even when two condensations are formed, they might approach each other and finally merge if the gas pressure between the two segments were low enough. Thus, we conjecture that the amplitude of the background heating $E_{0}$ also affects the subsequent evolution of threads. Keeping other parameters the same as in case RF1, we simulate different cases with $E_{0}=1.0,\ 3.0$ and $\ 4.0 \times \rm 10^{-4} \ erg\ cm^{-3}\ s^{-1}$, as listed in Table \ref{tab1}. In case BQ1 (Figure \ref{fig4}g), coronal condensations take place in both magnetic dips and finally coexist stably. However, in case BQ2, although condensations form in both dips initially, the two approach each other against the force of gravity and coalesce near the flux-tube midpoint under the inward pressure-gradient force; the resulting single condensation passes through the right-hand dip and finally falls to the chromosphere (Figure \ref{fig4}h). Thus, we suggest that the strong background heating is not conducive to the preservation of two threads. In particular, coronal condensations cannot be initiated when the background heating exceeds the radiative losses (e.g., case BQ3), as shown in Figures \ref{fig6}e and \ref{fig6}f. In addition, the background heating dominates over the other energy terms in this case. Recently, \citet{Terradas2021} found that no solutions with cold threads exist when the background heating is larger than radiative losses in a flux tube, which is consistent with our simulations. The scale length of the background heating is assumed to be much larger than that of the localized heating, indicating that the background heating decays more slowly with distance from the footpoints. Therefore, strong background heating may well dominate the fast cooling due to the radiation losses in the locations where condensations would form, so that thermal runaway cannot occur.

\subsubsection{Geometry of the magnetic flux tube}

Previous simulations of single-dipped flux tubes have demonstrated that the filament formation and dynamic processes depend significantly on the geometry of the flux tube \citep{Karpen2003}. Consequently, it is essential to understand the influences of the magnetic configuration on the formation of two threads in a single-dipped flux tube. 

The gravity distribution along the helical flux tube is fully controlled by $D$, $l$, and $n$. Hence, in the following simulations, we investigate the effects of these geometric parameters. To aid in interpreting the next set of simulations, we define three dimensionless parameters to describe the ability of a magnetic dip to host a thread. According to the field-aligned gravity along the flux tube, $g_{\parallel}$, the dip center is determined by the zero point in the monotonically decreasing interval of $g_{\parallel}$ along the helical portion, and the dip shoulders are determined by two associated zero points in the increasing interval, as depicted in Figure \ref{fig7} for case RF1. The points A, C, and B represent two dip shoulders, and the dip center, respectively, $S_{l}$ ($S_{r}$) is the length of left (right) portion of the dip, $S_{d}=S_{l}+S_{r}$ is the total length, and $g_{_{max} }$ ($g_{_{min} }$) is the maximum (minimum) field-aligned gravity inside the dip. Accordingly, the effective gravity of a dip $g_{_{eff} }$ is quantified by  the absolute value of the field-aligned gravity integrated over the whole dip: 
\begin{eqnarray}
 && g_{_{eff}} = \frac{\int_{0}^{S_{_{d}}} |g_{\parallel}(s)|ds} {S_{_{d}}g_{\odot}},
 \label{eq8}
\end{eqnarray}
The gravity asymmetry $A_{g}$ between the two sides of the dip is:
\begin{eqnarray}
 && A_{g}=\frac{min(|g_{_{max}}|,|g_{_{min}}|)}{max(|g_{_{max}}|,|g_{_{min}}|)},
 \label{eq9}
\end{eqnarray}
and the geometry asymmetry $A_{s}$ between two sides of the dip is determined by Equation (\ref{eq10}). 
\begin{eqnarray}
 && A_{s}=\frac{min(S_{l},S_{r})}{max(S_{l},S_{r})},
 \label{eq10}
\end{eqnarray}
Below we use these three dimensionless parameters to describe the ability of the dip to trap a filament thread, i.e., the trapping ability increases as the $g_{_{eff} }$ becomes larger, and $A_{g}$ and $A_{s}$ approach 1. Unlike the heating parameters explored in Section \ref{sec:helical_h}, the characteristics of the flux tube geometry are easier to deduce from observations. 

\begin{figure}[ht!]
\centering
\includegraphics[scale=0.65]{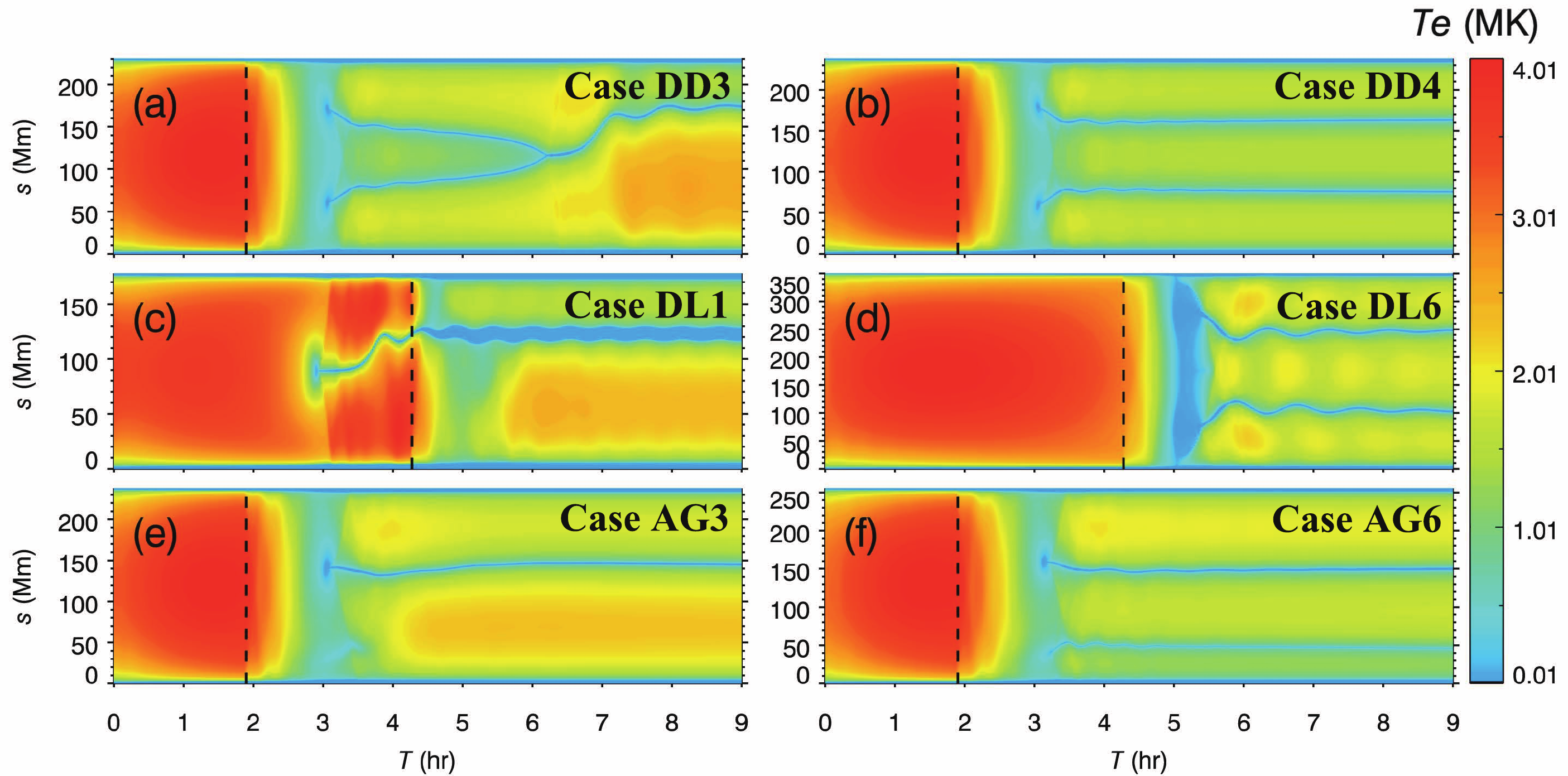}
\caption{The evolution of the temperature distribution along the flux tube in 6 cases with different flux-tube geometries. Three rows from the top to bottom show the influences of the dip depth $D$, the flux-tube length $l$, and the asymmetric metric $n$ on the formation of two threads, respectively. Panels (a)--(f) represent the results in cases DD3, DD4, DL1, DL6, AG3, AG6, respectively. The black dashed lines represent the time when the localized heating is halted.}
\label{fig8}
\end{figure}

To investigate the effects of varying geometrical factors on thread formation in our helical flux tube, we perform parameter studies of the dimensional flux-tube properties $D$, $L$, and $n$. Table \ref{tab2} lists the parameters and results, using the same notations for the results as in Table \ref{tab1}. The gravity distribution along a flux tube is highly associated with the depth of its magnetic dips. In this group of simulations (cases DD1--DD4), we select four values of $D$ = 3, 5, 7, and 11 Mm, with the localized heating parameters the same as in case RF1, as listed in Table \ref{tab2}. Note that case RF1, with $D=9$ Mm, fits between cases DD3 and DD4. We find that two threads can only exist stably on deeply dipped ($D$ = 9 Mm and 11 Mm) magnetic field lines. When $D<9\ \rm Mm$ ($g_{_{eff}}<0.19$), the gravity cannot overcome the gas pressure-gradient force, so the two threads finally approach each other and merge into one segment, e.g., case DD3, as shown in Figure \ref{fig8}a. On the contrary, when $D \geq 9\ \rm Mm$ ($g_{_{eff}}>0.19$), two condensations can form and persist in the two dips (Figure \ref{fig8}b). In case DD3, the thread temperature remains unchanged basically after the coalescence, whereas the thread density and length increase, as illustrated by Figures \ref{fig5}h and \ref{fig5}i. The power density distribution in case DD3 (Figure \ref{fig6}g) is similar to that in case RF1 (Figure \ref{fig6}a) at the onset time of the catastrophic cooling, indicating that the dip geometry is also important for the number of stationary threads trapped in dips.

To investigate the effects of the flux-tube length, we perform a series of simulations with $l$, ranging from 120 to 300 Mm, with other parameters being the same as in case RF2.  This range is within the observed scope of filament lengths \citep{Tanberg1995}. The reason of selecting the case RF2 as the benchmark is that two threads cannot form in this case, which can highlight the effects of the flux-tube length. Figures \ref{fig8}c and \ref{fig8}d depict the results in cases DL1 and DL6, respectively. As displayed in Table \ref{tab2}, only one condensation takes place when $l < 240\ \rm Mm$, while condensations form in both dips when $l>240\ \rm Mm$, for the following reasons. First, according to Equation (\ref{eq5}), increasing $l$ has a similar effect as decreasing $\lambda$. Second, we find that the onset time $\tau_{_{f}}$ increases with the flux-tube length, so that the $\tau_{_{f}}>\tau_{_{l}}$ condition can be met more easily, favoring the formation of two threads according to Section \ref{sec:helical_h}.

In the above two parameter surveys, the two magnetic dips are set to be identical widths, i.e., $n=1$ in Equation \ref{eq6}. However, on the Sun, the two dips along a single field line are probably different. To obtain different dip widths in the same flux tube, as in \citet{Zhou2017}, we adjust $n$ in Equation \ref{eq6}. If $n>1$, then $w_{2}>w_{1}$, i.e., the second (wider) dip is more extended. We take $n=1.1,\ 1.5,$ and $1.9$, with other parameters being the same as in case RF1. Only one segment of thread is formed in the second dip when $n>1.1$ (case AG3, Figure \ref{fig8}e), for the following reasons. First, the second dip is longer and can accumulate much more material. Second, the second dip center is further from the footpoints than the first dip. For example, in case AG3, the distances between each dip center and nearest footpoint are about 61 Mm and 100 Mm, respectively. Thus, the heating in the second dip is weaker, which facilitates the occurrence of thermal runaway. According to the above results for cases DD1--4, deep magnetic dips favor the formation of two threads. Therefore, in cases AG4--6, we fix $n=1.9$ and take $D=11,\ 13$ and 15 Mm. As expected, two condensations can also form if both dips are deep enough (Figure \ref{fig8}f).

\subsection{TDm magnetic flux rope model} 
\subsubsection{Thread formation in magnetic field lines with two dips}

\begin{figure}[ht!]
\centering
\includegraphics[scale=1.2]{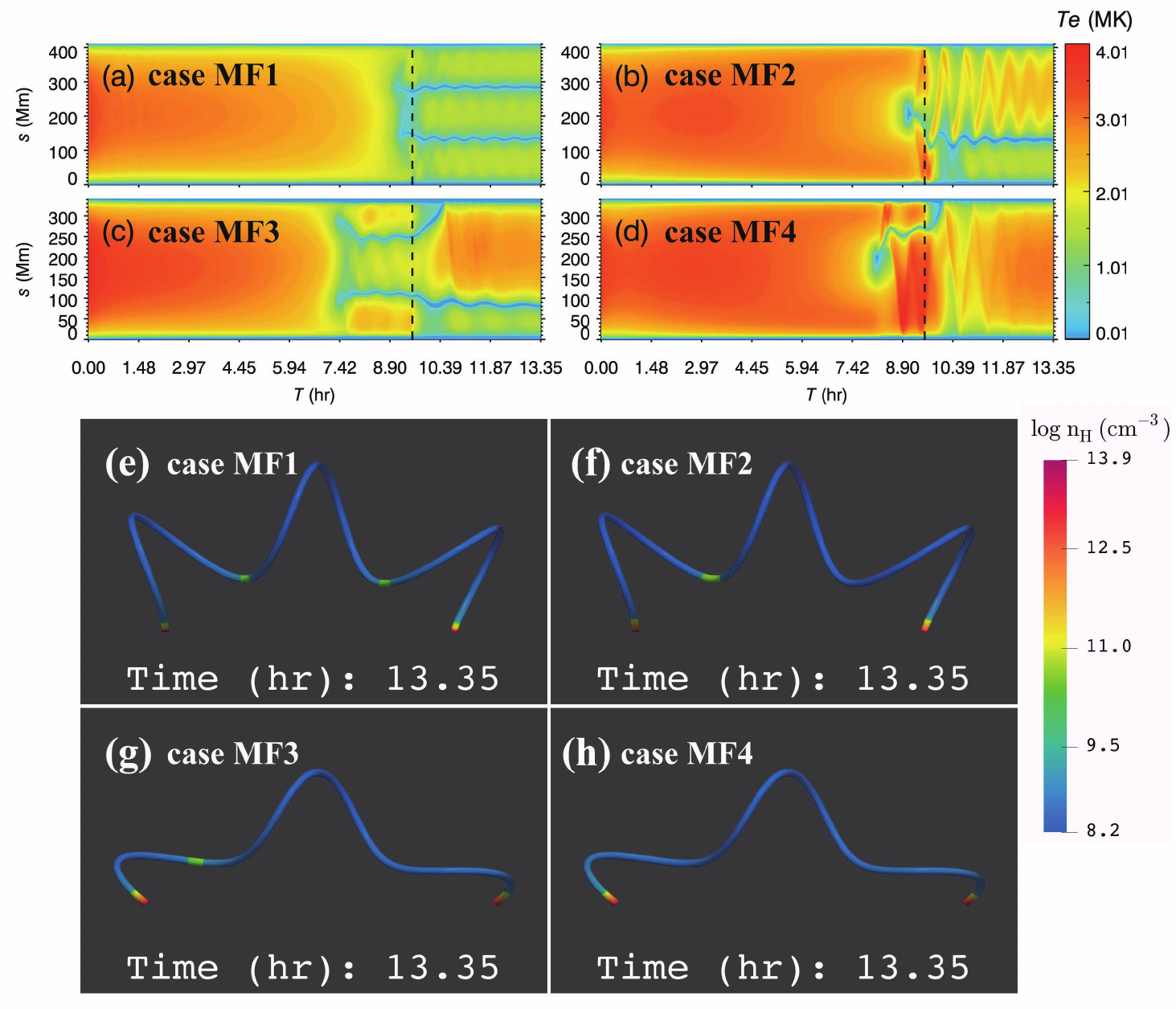}
\caption{Panels (a)--(d): The evolution of the temperature distribution along the flux tube in cases MF1, MF2, MF3 and MF4, respectively. The black dashed lines represent the time when the localized heating is halted. Panels (e)--(h): The number density distribution along the magnetic field line in cases MF1, MF2, MF3 and MF4 at the end of the runs, respectively.}
\label{fig9}
\end{figure}

In this section, we consider the condensation process in flux tubes extracted from a TDm force-free magnetic flux rope (see Section \ref{sec:met}). To investigate numerically the effects of the flux rope geometry, we select two different field lines hosting two dips, labeled MFL1 and MFL2, whose twists are 2.64 and 1.88 turns, and lengths are 411.4 and 342.6 Mm, respectively. We consider two localized heating scales: more concentrated ($\lambda=10\ \rm Mm$), and less concentrated localized heating ($\lambda=20\ \rm Mm$). In the simulations of this subsection, we take $\tau_{_{l}}= 573\ \rm min$ to evaporate enough chromospheric material for cold condensations to form on these long magnetic field lines. The input parameters and the simulation results are listed in Table \ref{tab3}. Figures \ref{fig9}a--\ref{fig9}d show the evolution of the temperature distributions along the flux tubes in cases MF1, MF2, MF3 and MF4, respectively. In Figures \ref{fig9}e--\ref{fig9}h, the final number density distributions along the 3D field lines are rendered in color.

We found that the localized heating scale is crucial to the occurrence of two cold condensations. When $\lambda = 10$ Mm, two condensations form near two dips (Figures \ref{fig9}a and \ref{fig9}e). By contrast, only one condensation forms at the loop center when $\lambda =20$ Mm (Figures \ref{fig9}b and \ref{fig9}f). In addition, the catastrophic coolings have taken place before the localized heating ended in all four cases. Therefore, we suggest that the localized heating scale is more significant than the heating duration. As long as the localized heating is concentrated enough, two condensations can be formed even under long-duration localized heating.

However, whether threads can survive also depends on the geometry of magnetic dips. For example, condensations can stably persist in magnetic dips in case MF1 (Figures \ref{fig9}a and \ref{fig9}e). On the contrary, one thread quickly drains into the chromosphere in cases MF3 and MF4 (Figures \ref{fig9}c, \ref{fig9}d, \ref{fig9}g and \ref{fig9}h). In order to explain this phenomenon, we calculate $g_{_{eff}},\ A_{g}$ and $A_{s}$ (Equations \ref{eq8}--\ref{eq10}) to describe the magnetic dip characteristics, whose results are listed in Table \ref{tab3}. We found that the second dip in MFL2 is highly asymmetric ($A_{g}=0.01,\ A_{s}=0.12$) so that the thread cannot persist in this dip. Thus, these three dimensionless parameters can well reflect the potential that the magnetic dip traps the filament thread, consistent with the effect of the slope of the dip \citep{Karpen2003}.

\subsubsection{Pseudo-3D simulation of filament formation in a magnetic flux rope}

As illustrated in Figure \ref{fig2}, the TDm flux rope core is weakly twisted while the outer shell is highly twisted. This magnetic model includes non-dipped, single-dipped and multiply-dipped field lines. Thus, the filament thread formation process and characteristics in different flux tubes would be significantly different even in a single flux rope. In this section, we explore the thread characteristics in a magnetic flux rope system with a pseudo-3D simulation, similar to \citet{luna2012}: many independent 1D simulations, in which the heating scale and duration are the same as in case MF1. To do so, we first select 40 flux tubes extracted from the TDm model, which are almost uniformly distributed inside the flux rope. Then, we omit 4 flux tubes with more than two dips, and perform 1D simulations in the remaining 36 flux tubes.

\begin{figure}[ht!]
\centering
\includegraphics[scale=0.65]{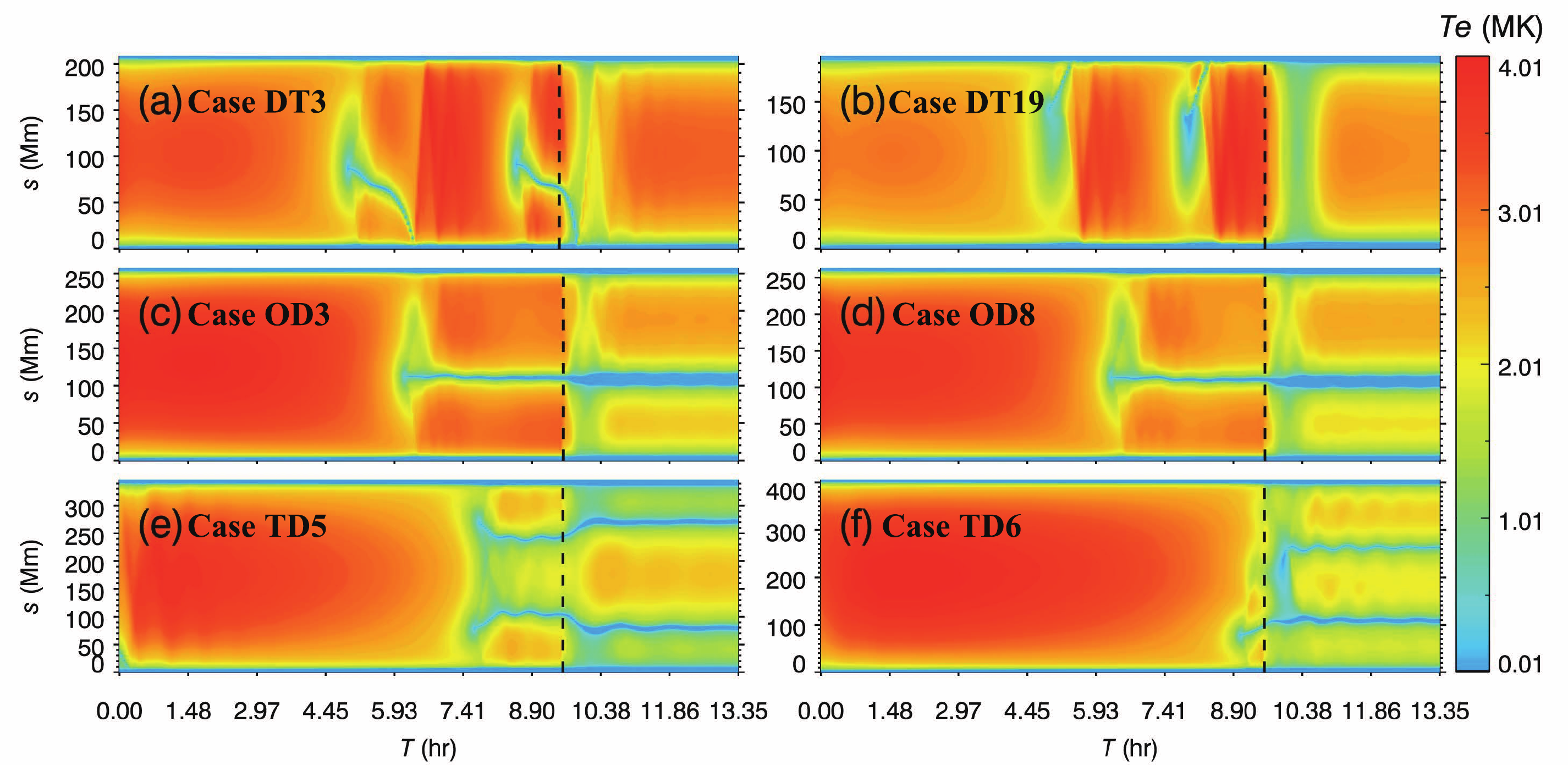}
\caption{The evolution of the temperature distribution in 6 of the selected flux tubes. Three rows from the top to bottom show the non-dipped (dynamic threads) cases, single-dipped (independently trapped threads) cases and double-dipped (magnetically connected threads) cases, respectively. Panels (a)--(f) represent the results in cases DT3, DT19, OD3, OD8, TD5, TD6, respectively. The black dashed lines represent the time when the localized heating is halted.}
\label{fig10}
\end{figure}

Table \ref{tab4} lists the topological characteristics of the selected magnetic field lines and the simulation results. The selected flux tubes can be classified into three groups: non-dipped cases ($55 \%$), single-dipped cases ($28 \%$), and double-dipped cases ($17 \%$). Note that these aforementioned fractions correspond to the magnetic model presented in this paper. Accordingly, there are three kinds of threads in a filament system: dynamic threads, independently trapped threads, and magnetically connected threads. Regarding the dynamic threads, where the magnetic dips are insignificant, threads flow rapidly in flux tubes \citep{Karpen2001}. Figure \ref{fig10} depicts the temporal evolution of the temperature distribution along the flux tube in 6 examples: three rows from the top to bottom show two dynamic thread, independently trapped thread, and magnetically connected thread cases, respectively.

\begin{figure}[ht!]
\centering
\includegraphics[scale=0.40]{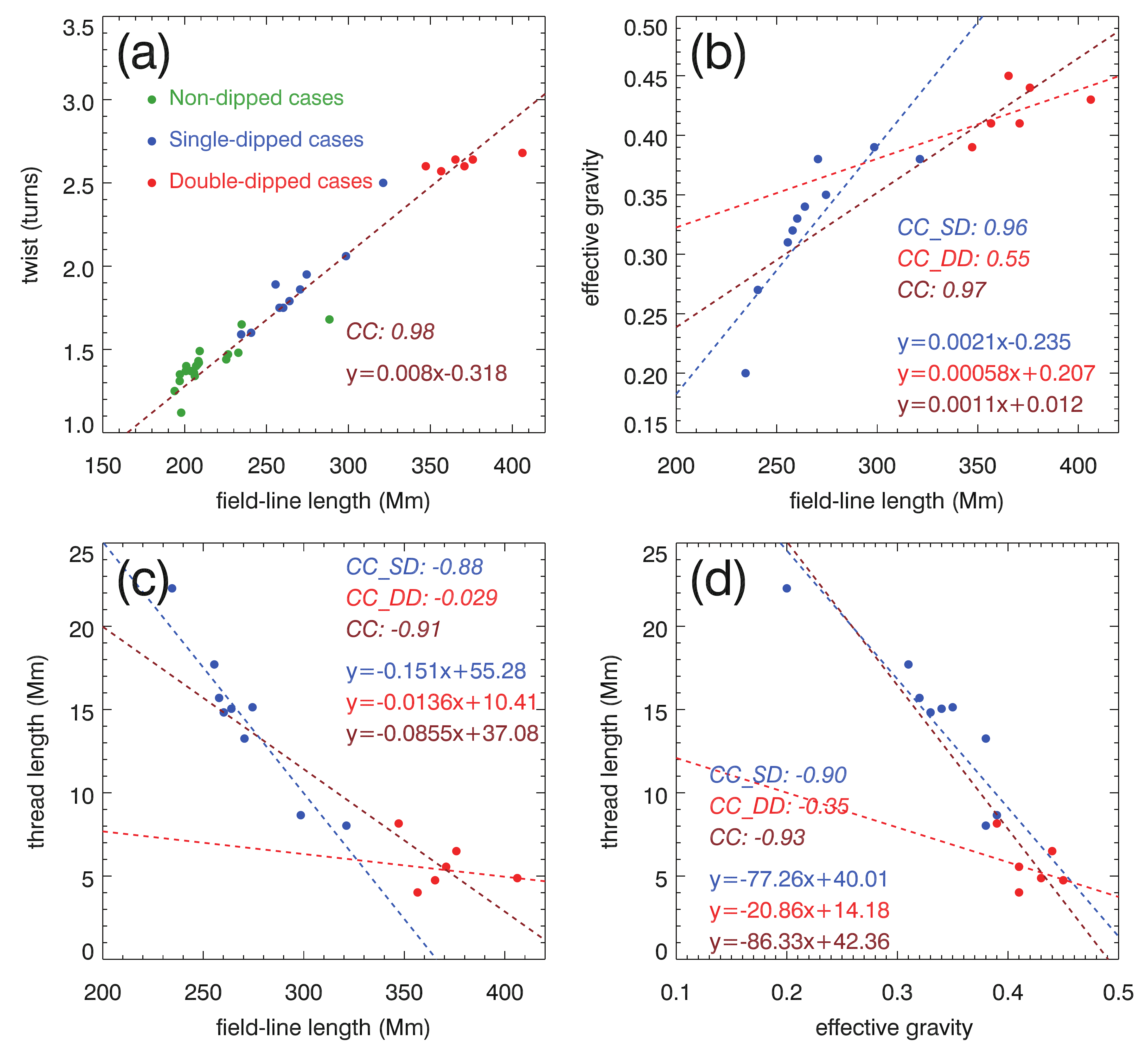}
\caption{Scatter plots of (a) twist vs field-line length; (b) effective gravity vs field-line length (maximum effective gravity for double-dipped field lines); (c) mean thread length vs field-line length; (d) mean thread length vs mean effective gravity on a field line. The green, blue, and red solid circles refer to non-dipped, single-dipped, and double-dipped cases, respectively. The dashed lines marked in blue, red and brown display the linear fitting results to single-dipped cases, double-dipped cases and all data points, respectively. CC\underline{~~}SD = correlation coefficient of single-dipped cases. CC\underline{~~}DD = correlation coefficient of double-dipped cases. CC = correlation coefficient of all data points.}
\label{fig11}
\end{figure}

Figure \ref{fig11}a shows the relationship between the lengths of the selected field lines and their twists. We find that the twist of a field line increases with its length with a correlation of 0.98, and the double-dipped field lines are commonly long and twisted. The effective gravity $g_{_{eff}}$ is also proportional to the field-line length with a correlation of 0.97 (Figure \ref{fig11}b), and increases more slowly in double-dipped cases. In Figures \ref{fig11}c and \ref{fig11}d, we find that the thread length is roughly inversely proportional to the field-line length and the mean effective gravity of dips on one field line, with correlations of -0.91 and -0.93, respectively. Moreover, the linear functions fitted from the double-dipped cases have smaller slopes and correlation coefficients than those of single-dipped cases. In particular, the threads on double-dipped field lines are usually shorter than those on single-dipped field lines. It is noted that these results are applicable to the TDm model; different flux-rope models might yield different correlations.

\begin{figure}[ht!]
\centering
\includegraphics[scale=0.45]{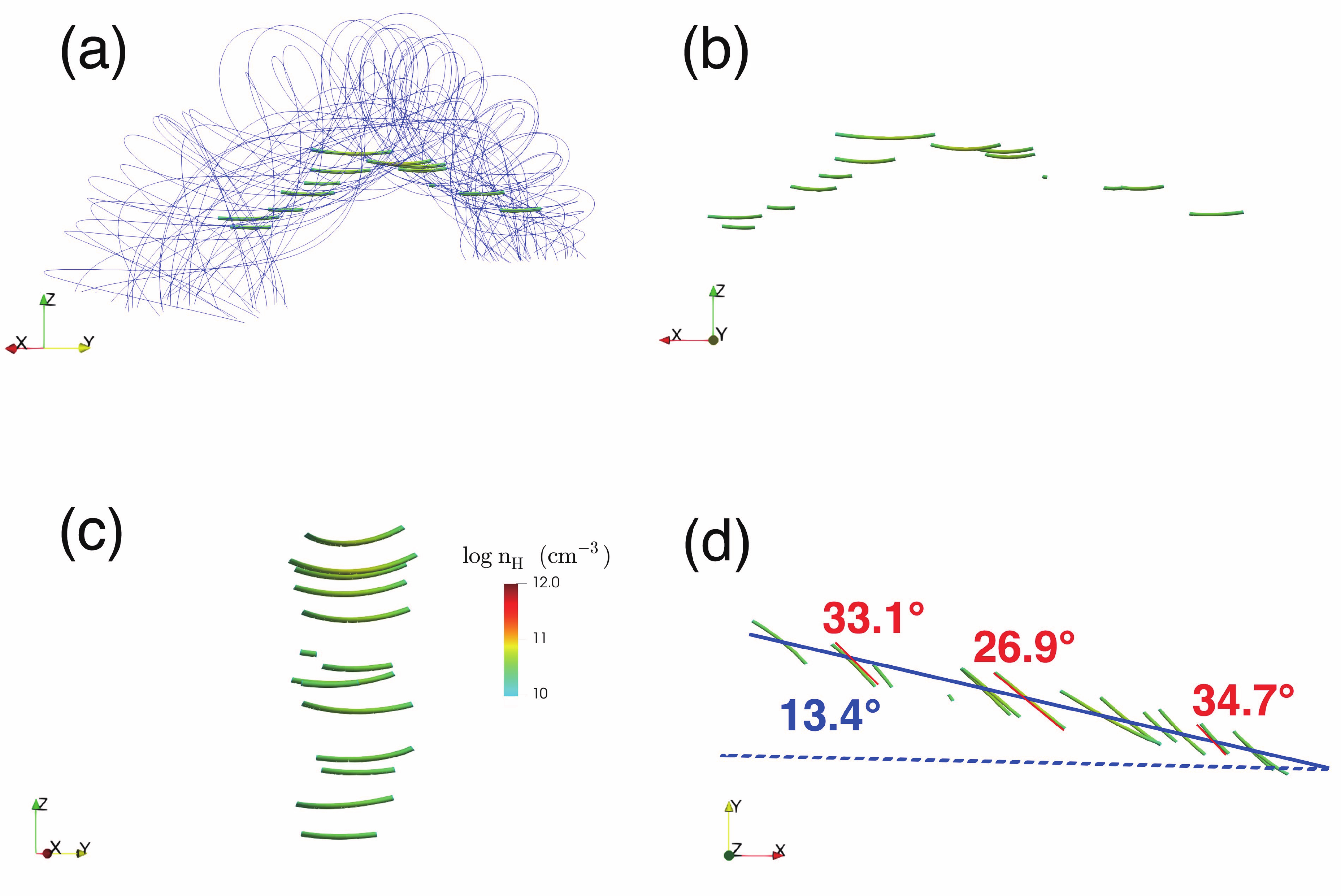}
\caption{Side and top views of the final thread distribution in 3D space. (a) Thread distribution on the magnetic field lines. The blue lines refer to magnetic field lines of the TDm flux rope. Panels (b)--(c): Side views of the thread distribution viewed along the $y$-axis and the filament axis, respectively. (d) Top view of the thread distribution, where the blue solid line denote the filament axis, the angle marked in blue represents that between the filament axis and the flux rope axis, and the angles marked in red represent those between the thread directions and the filament axis.} 
\label{fig12}
\end{figure}

Figure \ref{fig12} shows the side and top views of the filament thread distribution in 3D. We find that filament threads persist in the magnetic dips below the flux-rope axis as shown in Figure \ref{fig12}a. Figure \ref{fig12}b shows that this filament resembles an arcade with a bulge in the middle, consistent with many observed prominences. The filament is clearly composed of threads with different lengths. Figure \ref{fig12}c shows an end-on view of this  prominence composed of many vertically distributed horizontal threads. Figure \ref{fig12}d clearly displays that the filament threads are right-bearing, implying that the filament axial magnetic field is dextral in this flux rope and has negative helicity. Therefore the observed thread orientation can be used to determine the helicity of the underlying flux rope. Moreover, the angle between the filament axis and the flux rope axis is around $13^{\circ}$, which is similar to the 3D simulation of \citet{Xia2014}. Finally, we find that the pitch angles (between threads and the filament axis) vary from $25^{\circ}$ to $35^{\circ}$, within the range of the measured coronal magnetic-field pitch angle in a prominence \citep{Lopez2003}.

\section{Discussion} \label{sec:dis}
\subsection{The conditions to form magnetically connected threads}
In this paper, we focus on filament thread formation in a double-dipped flux tube. We find that two threads can exist stably in a double-dipped flux tube with specific parameters of the magnetic configuration and localized heating. The following conditions favor the formation of two magnetically connected threads: 1) Shorter localized heating scale; 2) Long magnetic flux tube with symmetric and deep dips.

According to the TNE theory, the thermal runaway starts at a place where the optically thin radiative losses exceed the heating, enthalpy and conductive fluxes. In this mechanism, chromospheric plasma is evaporated to the corona after a period of localized heating, increasing the coronal density. If the heating is concentrated near the flux-tube footpoints, the radiative losses, which are proportional to the density squared and increase with decreasing temperature $T$ for $T>100000$ K, could dominate the heating and other local energy sources far from the footpoints, cooling down the coronal plasma. Once the temperature decreases drastically, a low-pressure region is formed and then the surrounding plasma is sucked into this cold region, leading to a dense, cool condensation, namely, a filament thread. On the basis of our simulations and previous works \citep{Karpen2006, Xia2011}, whether two coronal condensations can occur is significantly affected by the spatial scale ($\lambda/L$), and the duration of the localized heating ($\tau_{_{l}}$), which dictate how much mass is evaporated into the flux tube and the occurrence of the thermal runaway. For long-duration ($\tau_{_{l} }>\tau_{_{f} }$) localized heating cases with large $\lambda/L$, the heating is too strong to allow the entire coronal portion of the flux tube to cool. Besides, the plasma is squeezed to the flux-tube center due to the large pressure gradient resulting from the strong footpoint heating, increasing the radiative losses there. Thus, only one condensation is formed at the flux-tube center and then falls into one dip, consistent with single-dipped cases with symmetric heating \citep{Karpen2006, Xia2011}. Meanwhile, with short-duration localized heating ($\tau_{_{l} }<\tau_{_{f} }$), the only heating is contributed by the background heating when the thermal runaway starts, so two condensations form at outer downslope of the dips after the cooling down of the entire coronal portion of the flux tube. Condensations can be trapped in magnetic dips that are deep and symmetric enough (i.e., case DD4). Thus, the short-duration localized heating enables the formation of magnetically connected threads on a double-dipped field line when the localized heating scale is large. However, if the localized heating scale is short enough, threads are formed in both magnetic dips even with long-duration localized heating (i.e., cases MF1, SL1--SL5, DL4--DL6). Since the localized heating and associated thermal conduction away from the heat source decay quickly with distance from the footpoints, the heating in the coronal portion of the flux tube is weak enough to allow two condensations to form and remain in dips.

TNE theory can tell us whether coronal condensation can occur. However, how many threads can be formed and persist also depends on the subsequent evolution. In some cases, even if the catastrophic cooling occurs in two locations, only one thread finally survives, e.g., cases SL3--SL5, and DD3. In these cases, two threads approach each other and finally merge into one segment due to the low-pressure region between two dips (Figure \ref{fig5}a), which is caused by the continued radiative losses and the accretion of filament threads. It is noted that, if $\lambda/L$ is too small, two threads are likely to fall back to the chromosphere (case SL1), as found in previous studies of a single-dipped flux tube \citep{Karpen2006,Xia2011}. Thus, we propose two qualitative conditions for two threads to form and persist in a double-dipped flux tube: 1) Thermal condition (appropriate $\lambda/L$ and $\tau_{_{l}}$): two coronal condensations can occur in the flux tube. Besides, in order to make sure that threads do not fall down to the chromosphere, the localized heating scale length, $\lambda$, should not be too short. 2) Dynamic condition (appropriate dip): two segments cannot coalesce under the inward pressure-gradient force. That is, the magnetic dip is able to trap the filament thread, i.e., larger $g_{_{eff}},\ A_{g}\ \sim 1,\ A_{s}\ \sim 1$. For asymmetric flux-tube dips, the relative position between the shallow side of a dip and the nearest footpoint also affects whether a condensation could be trapped in a dip. If the shallow side is further from the nearest footpoint, the condensation could be pushed toward the midpoint and merge with the other thread, whereas if the steeper slope is closer to the flux-tube center, the condensation is unlikely to escape (e.g., case MF1).

Earlier we noted that $10\%$ of the 40 selected flux tubes in the TDm model contained 3 dips. When a flux tube has more than 2 dips, we expect that the thread formation will be more complicated, with the possibilities of one, two or more threads being formed, depending on the magnetic configuration and the heating profile. A full parameter study of flux tubes with more than 2 dips is complicated by the increased number of free parameters, and thus is beyond the scope of this study. However, we simulated a symmetric flux tube with three dips by assuming symmetric heating at the two footpoints. We find that, if the scale height of the heating is small, three threads are formed, with the central one being only half as wide as the others; if the scale height is large, only one thread is formed in the central dip. Such tendencies are consistent with the results presented above. 

\subsection{The characteristics of magnetically connected threads and the filament fine structure}
\begin{figure}[ht!]
\centering
\includegraphics[scale=0.8]{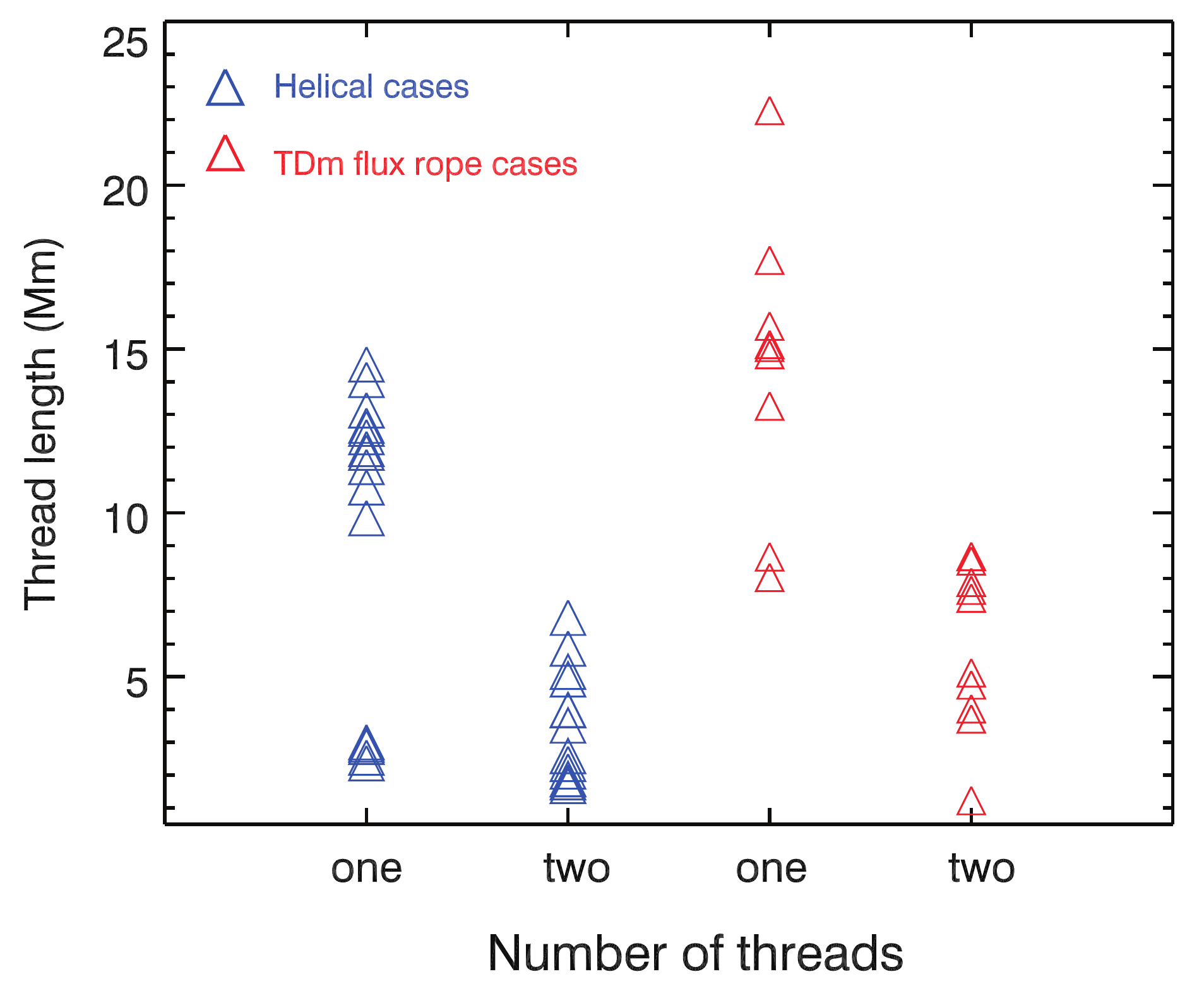}
\caption{Scatter plot of the thread length versus the number of threads. Blue and red triangles refer to helical cases and TDm flux rope cases, respectively.}
\label{fig13}
\end{figure}

The thread length is an important property of solar filaments, which can reflect the magnetic configuration of the supporting field lines \citep{Karpen2003, Zhou2014}. To investigate the relationship between the number of threads and the thread length, we make a statistical analysis of all helical cases, as shown by blue triangles in Figure \ref{fig13}. For the cases with two threads, the maximum thread length is 6.8 Mm, the minimum length is 0.34 Mm, and the average length is 2.92 Mm. For the cases with only one thread, the maximum thread length is 14.53 Mm, the minimum length is 2.36 Mm, and the average length is 9.66 Mm. However, the flux tubes are independent in our helical cases. For a set of flux tubes extracted from the TDm flux-rope system shown in Figure \ref{fig2} (red triangles in Figure \ref{fig13}), we find that the average length of the independently trapped threads is about 14.52 Mm, but that of the magnetically connected threads is about 5.89 Mm. Thus, in a statistical sense, the length of each thread is highly correlated with the number of threads along one flux tube. The average lengths differ by close to a factor of 3, as expected: a magnetically connected thread is significantly shorter on average than an independently trapped thread. The primary reason is that a finite amount of mass is evaporated for a given heating rate and scale height, which is shared by both threads.

Many studies have shown that active-region filaments and quiescent filaments differ in many aspects, such as their height, length, dynamics, magnetic-field strength, and magnetic configuration \citep{Tanberg-Hanssen1974, Tanberg1995, Filippov2000, Moore2000, Kuckein2009, Mackay2010, Engvold2015, Xing18}. According to \citet{Ouyang2017}, $96 \%$ of quiescent filaments are supported by flux ropes. By contrast, only $60 \%$ of active-region filaments are supported by flux ropes. It is conceivable that the dips of flux ropes are generally deeper than those of sheared arcades, particularly in the outer portions of the flux rope, and that at least some of the field lines of long quiescent filaments have more than one dip. In general, quiescent filaments are longer than active-region filaments, which implies that quiescent filaments are likely to have longer field lines. Additionally, quiescent filaments have weaker magnetic fields than active-region filaments, implying that the background heating could be weaker for quiescent filaments than for active-region filaments. Both features favor the formation of magnetically connected threads in the flux tubes of quiescent filaments. More intriguingly, high-resolution observations revealed that quiescent filaments are usually composed of relatively shorter threads than active-region filaments \citep{Lin2008a, Mackay2010}. Our simulation results indicate that the thread length is significantly shorter in cases with two threads than in cases with only one segment of thread, which might provide a natural explanation of shorter threads in quiescent filaments.

In addition to the thread length, quiescent filament fine structures have some puzzling features, such as barbs and vertical threads. Observations indicate that quiescent filaments have more barbs than active-region filaments \citep{Hao2015, Chen2020}. \citet{Chen2020} summarized different types of filament barbs, i.e., (1) barbs corresponding to the feet extending down to the solar surface \citep{Aulanier1998}; (2) dynamic barbs due to the longitudinal oscillations of some threads \citep{Ouyang2020}; (3) barbs due to the indented threads. Note that the latter two kinds of barbs do not correspond to any prominence footpoint. According to our simulations, magnetically connected threads are usually shorter than independently trapped threads in a filament system whose structure is consistent with a TDm flux rope. If the threads are not aligned with the filament axis (Figure \ref{fig12}d), the combinations of long, independently trapped threads and short, magnetically connected threads easily resemble barbs, so we suggest that barbs are likely to be present in a long quiescent filament supported by a twisted flux rope.

In general, filament threads are nearly parallel with the filament-channel PIL. However, some quiescent prominences, especially polar crown prominences, show vertical threads \citep{Berge2008, Su2012}. It is still a puzzle why these prominence threads are vertical. Based on a 3D simulation, \citet{Xia2016b} proposed that the magnetic Rayleigh-Taylor instability can influence the dynamics inside the prominence, forming the vertical threads. Other researchers argued that the vertical threads in quiescent prominences are due to the piling up of many small dips containing short threads \citep{Schmieder2014, Ruan2018}. Our simulation results are consistent with the latter view.

\section{Summary} \label{sec:sum}

Previous 1D simulations of single-dipped field lines have demonstrated that the evaporation--condensation model can explain many observational features of filament mass formation \citep{Antiochos1999, Karpen2003, Karpen2005, Karpen2006, Karpen2008, Xia2011, Zhou2014}. However, observations and theories indicate that some of the coronal magnetic-field lines in filament channels might possess more than one dip \citep{Jing2010, Martens2001, Mackay2010, Su2015, Guo2019}. We expect that the thread formation on a field line with two dips could be much different than found in previous studies with one dip. For example, if two dips exist along the same field line and contain cool plasma, thread-thread interactions could occur \citep{Zhou2017, Zhang2017}. To explore filament thread formation in a double-dipped magnetic flux tube, we performed 1D hydrodynamic simulations with optically thin radiative cooling, varying the flux-tube geometry and key heating parameters. The main results are summarized as follows.

\begin{enumerate}
\item{In a double-dipped flux tube, threads do not necessarily form or persist in both dips. In many cases, only one thread is formed, which tends to stay in the longer, deeper and symmetric dip near the more weakly heated footpoint. Two magnetically connected threads tend to be formed under the following conditions: for the thermal condition, shorter localized concentrated heating scale can make thermal runaway occur in both magnetic dips rather in the loop center; for the dynamic condition, magnetically connected threads persist more easily in deep (larger $g_{_{eff}}$) and symmetric magnetic dips ($A_{g}\ \sim 1, A_{s}\ \sim 1$)}.

\item{According to \citet{Ouyang2017}, short active-region filaments tend to be associated with sheared magnetic field, whereas long quiescent filaments tend to be associated with twisted magnetic field, some of which are twisted more than 2 turns \citep[e.g.,][]{Su2015, Guo2019, Mackay2020}, forming more than 2 dips on some of the field lines. Based on our simulations, the condensations in the cases with two magnetically connected threads are significantly shorter on average than those in the cases with one condensation. Our numerical results imply that quiescent filaments would possess shorter threads than active-region filaments, as observed. Moreover, the short lengths of magnetically connected threads, as well as their locations in the highly twisted regions of the flux rope, might explain the formation mechanism of barbs and apparently vertical threads in quiescent filaments/prominences.}
\end{enumerate}

\acknowledgments
We thank ISSI and ISSI-Beijing for supporting team meetings on longitudinal oscillations of solar filaments. The numerical simulations in this paper were performed in the cluster system of the High Performance Computing Center of Nanjing University. This research was supported by NSFC (11961131002, 11533005, and 11773016), National Key Research and Development Program of China (2020YFC2201201), and the Science and Technology Development Fund of Macau, China (275/2017/A). Y.H.Z. is supported by the Belgian FWO-NSFC project G0E9619N.

\bibliography{ms}{}
\bibliographystyle{aasjournal}

\newpage
\begin{deluxetable*}{lccccccccccl}
\caption{Parameters and Results in Cases with Different Heating Characteristics \label{tab1}}
\tablewidth{0pt}
\tablehead{
\colhead{} & \colhead{Case} & \colhead{$\tau_{_{l}}$} & \colhead{$\lambda$} &
\colhead{$f$} & \colhead{$E_{0}$} & \colhead{$N_{_{threads} }$$^{\rm a}$} & \colhead{$\tau_{_{forml} }$$^{\rm b}$} & \colhead{$\tau_{_{formr} }$$^{\rm c}$} & \colhead{$L_{_{tl}}$$^{_{\rm d}}$} & \colhead{$L_{_{tr}}$$^{\rm e}$} \\
\colhead{} & \colhead{} & \colhead{(min)}  & \colhead{(Mm)} &
\colhead{} & \colhead{($\rm \times 10^{-4} \ erg\,cm^{-3}\,s^{-1}$)} & \colhead{} & \colhead{(min)} & \colhead{(min)} & \colhead{(Mm)} & \colhead{(Mm)}
}
\startdata
{}& RF1 & 115 & 10 & 1.0 & 2.0 & 2 & 184 & 184 & 1.88 & 1.88 \\
{}& HT2 & 143 & 10 & 1.0 & 2.0 & 2 & 192 & 192 & 2.58 & 2.58 \\
{$\tau_{_{l} }$}& HT3 & 172 & 10 & 1.0 & 2.0 & 2 & 204 & 204 & 3.99 & 3.98 \\
{}& HT4 & 201 & 10 & 1.0 & 2.0 & 2 & 212 & 212 & 5.86 & 6.80 \\
{}& HT5 & 229 & 10 & 1.0 & 2.0 & 1 & - & 216 & - & 11.95 \\
{}& RF2 & 258 & 10 & 1.0 & 2.0 & 1 & - & 218 & - & 12.66 \\
\hline
{} & SL1 & 258 & 3 & 1.0 & 2.0 & 0 & 248 & 248 & - & - \\
{} & SL2 & 258 & 4 & 1.0 & 2.0 & 2 & 209 & 209 & 4.92 & 4.92 \\
{$\lambda$} & SL3 & 258 & 5 & 1.0 & 2.0 & 1 & 199 & 199 & 9.84 & - \\
{}  & SL4 & 258 & 7 & 1.0 & 2.0 & 1 & 201 & 201 & 10.78 & - \\
{}  & SL5 & 258 & 9 & 1.0 & 2.0 & 1 & 214 & 214 & 12.66 & - \\
\hline
{} & AH1 & 115 & 10 & 1.1 & 2.0 & 2 & 182 & 182 & 1.88 & 1.88 \\
{} & AH2 & 115 & 10 & 1.4 & 2.0 & 2 & 179 & 179 & 1.87 & 2.11 \\
{$f$} & AH3 & 115 & 10 & 1.7 & 2.0 & 2 & 176 & 176 & 1.78 & 2.34 \\
{}  & AH4 & 258 & 10 & 1.1 & 2.0 & 1 & 219 & - & 13.13 & - \\
{}  & AH5 & 258 & 10 & 1.4 & 2.0 & 1 & 214 & - & 14.06 & - \\
{}  & AH6 & 258 & 10 & 1.7 & 2.0 & 1 & 211 & - & 14.53 & - \\
\hline
{} & BQ1 & 115 & 10 & 1.0 & 1.0 & 2 & 161 & 161 & 3.98 & 3.98 \\
{$E_{0}$}& RF1 & 115 & 10 & 1.0 & 2.0 & 2 & 184 & 184 & 1.88 & 1.88 \\
{}& BQ2 & 115 & 10 & 1.0 & 3.0 & 0 & 227 & 227 & - & - \\
{}& BQ3 & 115 & 10 & 1.0 & 4.0 & 0 & - & - & - & - \\
\enddata
\tablecomments{$^{\rm a}$ Number of threads at the end of the simulation.\\ $^{\rm b}$ The onset time of the catastrophic cooling in the left-side dip (near $s=0$). We define the onset time as when temperature decreases drastically within 1 min.\\ $^{\rm c}$ The onset time of the catastrophic cooling in the right-side dip (near $s=s_{_{max}}$).\\ $^{\rm d}$ The length of the thread in the left-side dip, which is determined by the coronal region below 20000 K.\\ $^{\rm e}$ The length of the thread in the right-side dip.}
\end{deluxetable*}
\newpage
\begin{deluxetable*}{lcccccccccccccccl}
\caption{Parameters and Results in Cases with Different Geometries} \label{tab2}
\tablewidth{0pt}
\tablehead{
\colhead{} & \colhead{Case} & \colhead{$D$} & \colhead{$l$} & \colhead{$n$} & \colhead{$g_{_{eff1}}$$^{\rm a}$} &
\colhead{$g_{_{eff2}}$} & \colhead{$A_{g1}$$^{\rm b}$} & \colhead{$A_{g2}$} & \colhead{$A_{s1}$$^{\rm c}$} & \colhead{$A_{s2}$} & \colhead{$\tau_{_{l} }$} & \colhead{$N_{_{threads}}$} & \colhead{$\tau_{_{forml}}$} & \colhead{$\tau_{_{formr}}$} & \colhead{$L_{_{tl}}$} & \colhead{$L_{_{tr}}$} \\
\colhead{} & \colhead{} & \colhead{(Mm)} & \colhead{(Mm)} & \colhead{ } & \colhead{} &
\colhead{} & \colhead{ } & \colhead{ } & \colhead{ } & \colhead{ }  & \colhead{(min)} & \colhead{ } & \colhead{(min)} & \colhead{(min)} & \colhead{(Mm)} & \colhead{(Mm)}
}
\startdata
{} & DD1 & 3 & 180 & 1.0 & 0.07 & 0.07 & 1.0 & 1.0 & 1.0 & 1.0 & 115 & 0 & 192 & 192 & - & - \\
{} & DD2 & 5 & 180 & 1.0 & 0.11 & 0.11 & 1.0 & 1.0 & 1.0 & 1.0 & 115 & 0 & 189 & 189 & - & - \\
{$D$} & DD3 & 7 & 180 & 1.0 & 0.15 & 0.15 & 1.0 & 1.0 & 1.0 & 1.0 & 115 & 1 & 185 & 185 & - & 3.00 \\
{} & RF1 & 9 & 180 & 1.0 & 0.19 & 0.19 & 1.0 & 1.0 & 1.0 & 1.0 & 115 & 2 & 184 & 184 & 1.88 & 1.88 \\
{} & DD4 & 11 & 180 & 1.0 & 0.23 & 0.23 & 1.0 & 1.0 & 1.0 & 1.0 & 115 & 2 & 184 & 184 & 1.67 & 1.67 \\
\hline
{} & DL1 & 9 & 120 & 1.0 & 0.27 & 0.27 & 1.0 & 1.0 & 1.0 & 1.0 & 258 & 1 & - & 176 & - & 12.31 \\
{} & DL2 & 9 & 150 & 1.0 & 0.23 & 0.23 & 1.0 & 1.0 & 1.0 & 1.0 & 258 & 1 & - & 198 & - & 12.57 \\
{} & RF2 & 9 & 180 & 1.0 & 0.19 & 0.19 & 1.0 & 1.0 & 1.0 & 1.0 & 258 & 1 & - & 218 & - & 12.66 \\
{$l$}  & DL3 & 9 & 210 & 1.0 & 0.17 & 0.17 & 1.0 & 1.0 & 1.0 & 1.0 & 258 & 1 & - & 242 & - & 11.84 \\
{} & DL4 & 9 & 240 & 1.0 & 0.15 & 0.15 & 1.0 & 1.0 & 1.0 & 1.0 & 258 & 1 & 264 & 264 & - & 11.40 \\
{} & DL5 & 9 & 270 & 1.0 & 0.13 & 0.13 & 1.0 & 1.0 & 1.0 & 1.0 & 258 & 2 & 285 & 285 & 5.15 & 5.15 \\
{} & DL6 & 9 & 300 & 1.0 & 0.12 & 0.12 & 1.0 & 1.0 & 1.0 & 1.0 & 258 & 2 & 304 & 304 & 3.51 & 3.51 \\
\hline
{} & AG1 & 9 & 180 & 1.1 & 0.20 & 0.18 & 0.90 & 0.97 & 0.88 & 0.97 & 115 & 2 & 186 & 186 & 1.17 & 2.11 \\
{} & AG2 & 9 & 180 & 1.5 & 0.26 & 0.15 & 0.62 & 0.85 & 0.62 & 0.85 & 115 & 1 & - & 185 & - & 2.36 \\
{$n$} & AG3 & 9 & 180 & 1.9 & 0.31 & 0.13 & 0.50 & 0.75 & 0.52 & 0.75 & 115 & 1 & - & 185 & - & 2.86 \\
{} & AG4 & 11 & 180 & 1.9 & 0.36 & 0.16 & 0.55 & 0.75 & 0.56 & 0.75 & 115 & 1 & - & 185 & - & 2.93 \\
{} & AG5 & 13 & 180 & 1.9 & 0.40 & 0.19 & 0.59 & 0.76 & 0.61 & 0.76 & 115 & 1 & - & 186 & - & 2.53 \\
{} & AG6 & 15 & 180 & 1.9 & 0.43 & 0.21 & 0.64 & 0.76 & 0.65 & 0.77 & 115 & 2 & 196 & 189 & 0.34 & 2.33 \\
\hline
\enddata
\tablecomments{$^{\rm a}$Effective gravity (Equation \ref{eq8}).\\ $^{\rm b}$Gravity asymmetry (Equation \ref{eq9}).\\ $^{\rm c}$Geometry asymmetry (Equation \ref{eq10}).\\The notations are the same as Table \ref{tab1}.}
\end{deluxetable*}

\newpage
\begin{deluxetable*}{lccccccccccccccl}
\caption{Parameters and Results of Simulations of Selected Double-Dipped TDm Flux-Rope Field Lines \label{tab3}}
\tablewidth{0pt}
\tablehead{
\colhead{} & \colhead{Case} & \colhead{$|T_{w}|$$^{a}$} & \colhead{$\tau_{_{l}}$} & \colhead{$\lambda$} & \colhead{$g_{_{eff1} }$} & \colhead{$g_{_{eff2} }$} & \colhead{$A_{_{g1} }$} & \colhead{$A_{_{g2} }$} & \colhead{$A_{_{s1} }$} & \colhead{$A_{_{s2} }$} & \colhead{$N_{_{threads} }$} & \colhead{$\tau_{_{forml} }$} & \colhead{$\tau_{_{formr} }$} & \colhead{$L_{_{tl} }$} & \colhead{$L_{_{tr} }$} \\
\colhead{} & \colhead{} & \colhead{} & \colhead{(min)} & \colhead{(Mm)} &
\colhead{} & \colhead{} & \colhead{} & \colhead{} & \colhead{} & \colhead{} & \colhead{} & \colhead{(min)} & \colhead{(min)} & \colhead{(Mm)} & \colhead{(Mm)}
}
\startdata
{MFL1} & MF1 & 2.64 & 573 & 10 & 0.51 & 0.49 & 0.61 & 0.55 & 0.85 & 0.83 & 2 & 556 & 556 & 4.53 & 5.35 \\
{} & MF2 & 2.64 & 573 & 20 & 0.51 & 0.49 & 0.61 & 0.55 & 0.85 & 0.83 & 1 & 543 & - & 8.23 & - \\
\hline
{MFL2} & MF3 & 1.88 & 573 & 10 & 0.38 & 0.47 & 0.17 & 0.01 & 0.48 & 0.12 & 1 & 449 & 449 & 6.47 & - \\
{} & MF4 & 1.88 & 573 & 20 & 0.38 & 0.47 & 0.17 & 0.01 & 0.48 & 0.12 & 0 & 486 & - & - & -\\
\enddata
\tablecomments{The notations are the same as Table \ref{tab2} but the geometry of the flux tube is represented by the twist.\\ $^{a}$ $|T_{\rm w}|$ denotes the absolute value of the twist of the magnetic field line.}
\end{deluxetable*}

\newpage
\begin{longrotatetable}
\begin{deluxetable*}{lcccccccccccccl}
\tablecaption{Parameters and Results of Simulations of 36 Selected TDm Flux-Rope Field Lines \label{tab4}}
\tablewidth{700pt}
\tabletypesize{\scriptsize}
\tablehead{
\colhead{} & \colhead{Case} & \colhead{$L$} & \colhead{$|T_{w}|$} & \colhead{$N_{_{dips}}$} & \colhead{$g_{_{eff1} }$} & \colhead{$g_{_{eff2}}$} & \colhead{$A_{g1}$} & \colhead{$A_{g2}$} & \colhead{$A_{s1}$} & \colhead{$A_{s2}$} & \colhead{$N_{_{threads}}$} & \colhead{$L_{_{tl}}$} & \colhead{$L_{_{tr}}$} \\
\colhead{} & \colhead{} & \colhead{(Mm)} & \colhead{} & \colhead{} &
\colhead{} & \colhead{} & \colhead{} & \colhead{} & \colhead{} & \colhead{} & \colhead{} & \colhead{(Mm)} & \colhead{(Mm)}
}
\startdata
{} & DT1 & 197.9 & 1.12 & 0 & - & - & - & - & - & - & 0 & - & - \\
{} & DT2 & 209.2 & 1.49 & 0 & - & - & - & - & - & - & 0 & - & - \\
{} & DT3 & 208.7 & 1.42 & 0 & - & - & - & - & - & - & 0 & - & - \\
{} & DT4 & 208.4 & 1.43 & 0 & - & - & - & - & - & - & 0 & - & - \\
{} & DT5 & 200.7 & 1.37 & 0 & - & - & - & - & - & - & 0 & - & - \\
{} & DT6 & 205.7 & 1.36 & 0 & - & - & - & - & - & - & 0 & - & - \\
{} & DT7 & 193.9 & 1.25 & 0 & - & - & - & - & - & - & 0 & - & - \\
{} & DT8 & 207.9 & 1.41 & 0 & - & - & - & - & - & - & 0 & - & - \\
{non-dipped} & DT9 & 194.0 & 0.80 & 0 & - & - & - & - & - & - & 0 & - & - \\
{} & DT10 & 201.0 & 1.40 & 0 & - & - & - & - & - & - & 0 & - & - \\
{} & DT11 & 225.4 & 1.44 & 0 & - & - & - & - & - & - & 0 & - & - \\
{} & DT12 & 203.8 & 1.37 & 0 & - & - & - & - & - & - & 0 & - & - \\
{} & DT13 & 232.8 & 1.48 & 0 & - & - & - & - & - & - & 0 & - & - \\
{} & DT14 & 226.5 & 1.47 & 0 & - & - & - & - & - & - & 0 & - & - \\
{} & DT15 & 206.2 & 1.34 & 0 & - & - & - & - & - & - & 0 & - & - \\
{} & DT16 & 288.4 & 1.68 & 0 & - & - & - & - & - & - & 0 & - & - \\
{} & DT17 & 207.0 & 1.40 & 0 & - & - & - & - & - & - & 0 & - & - \\
{} & DT18 & 234.8 & 1.65 & 0 & - & - & - & - & - & - & 0 & - & - \\
{} & DT19 & 197.0 & 1.31 & 0 & - & - & - & - & - & - & 0 & - & - \\
{} & DT20 & 197.1 & 1.35 & 0 & - & - & - & - & - & - & 0 & - & - \\
\hline
{} & OD1 & 270.5 & 1.86 & 1 & 0.38 & - & 0.84 & - & 0.92 & - & 1 & 13.26 & - \\
{} & OD2 & 255.5 & 1.89 & 1 & 0.31 & - & 0.92 & - & 0.94 & - & 1 & 17.71 & - \\
{} & OD3 & 257.9 & 1.75 & 1 & 0.32 & - & 0.51 & - & 0.94 & - & 1 & 15.70 & - \\
{single-dipped} & OD4 & 298.5 & 2.06 & 1 & 0.39 & - & 0.44 & - & 0.73 & - & 1 & 8.66 & - \\
{} & OD5 & 274.5 & 1.95 & 1 & 0.35 & - & 0.30 & - & 0.60 & - & 1 & 15.14 & - \\
{} & OD6 & 260.2 & 1.75 & 1 & 0.33 & - & 0.85 & - & 0.60 & - & 1 & 14.83 & - \\
{} & OD7 & 321.2 & 2.50 & 1 & 0.38 & - & 0.24 & - & 0.56 & - & 1 & 8.03 & - \\
{} & OD8 & 264.0 & 1.79 & 1 & 0.34 & - & 0.48 & - & 0.73 & - & 1 & 15.05 & - \\
{} & OD9 & 234.5 & 1.59 & 1 & 0.20 & - & 0.41 & - & 0.63 & - & 1 & 22.28 & - \\
{} & OD10 & 240.6 & 1.60 & 1 & 0.27 & - & 0.11 & - & 0.30 & - & 0 & - & - \\
\hline
{} & TD1 & 365.3 & 2.64 & 2 & 0.42 & 0.45 & 0.73 & 0.41 & 0.08 & 0.36 & 1 & - & 4.75 \\
{} & TD2 & 375.9 & 2.64 & 2 & 0.43 & 0.44 & 0.74 & 0.41 & 0.22 & 0.59 & 2 & 7.88 & 5.12 \\
{double-dipped} & TD3 & 370.8 & 2.60 & 2 & 0.41 & 0.17 & 0.52 & 0.43 & 0.43 & 0.74 & 2 & 3.71 & 7.40 \\
{} & TD4 & 365.6 & 2.57 & 2 & 0.41 & 0.09 & 0.41 & 0.42 & 0.43 & 0.71 & 1 & 4.02 & - \\
{} & TD5 & 347.2 & 2.60 & 2 & 0.39 & 0.11 & 0.40 & 0.39 & 0.09 & 0.37 & 2 & 8.67 & 7.64 \\
{} & TD6 & 406.2 & 2.68 & 2 & 0.43 & 0.28 & 0.67 & 0.50 & 0.75 & 0.91 & 2 & 8.53 & 1.22 \\
\enddata
\end{deluxetable*}
\tablecomments{The notations are the same as Table \ref{tab3}.}
\end{longrotatetable}

\newpage

\newpage

\newpage

\newpage

\newpage

\newpage

\newpage

\newpage

\newpage

\newpage

\newpage

\newpage

\end{document}